\documentclass[12pt]{article}
\usepackage{graphicx}
\usepackage{amsmath}
\usepackage{mathrsfs}
\usepackage[square, authoryear, sort, comma]{natbib}
\usepackage{bm}
\usepackage{subfigure}
\usepackage[rightcaption]{sidecap} 
\usepackage{wrapfig}
\usepackage{float}

\bibpunct[,]{[}{]}{,}{a}{,}{,}

\newcommand{\beq}{\begin{equation}}
\newcommand{\eeq}{\end{equation}}
\newcommand{\bea}{\begin{eqnarray}}
\newcommand{\eea}{\end{eqnarray}}

\begin{document}
\vspace{-10mm}
\begin{center} {\small  To appear in {\it The Oxford Handbook on Nanoscience and
Nanotechnology:\\ Frontiers and Advances, eds. A.V. Narlikar and
Y.Y. Fu, Vol. I, Ch. 21}}\\ \vspace*{5mm} { Sergei Sergeenkov \\
} \vspace*{2mm} { Departamento de F\'isica, Universidade Federal
da Para\'iba, Jo\~ao Pessoa, Brazil}
\end{center}
\vspace*{5mm} \begin{center}{\large \bf 2D ARRAYS OF JOSEPHSON
NANOCONTACTS\\ \vspace*{1mm} AND NANOGRANULAR SUPERCONDUCTORS}\\
\vspace*{5mm} ABSTRACT\\
\end{center}  \vspace*{2mm}
{\small By introducing a realistic model of nanogranular
superconductors (NGS) based on 2D arrays of Josephson nanocontacts
(created by a network of twin-boundary dislocations with strain
fields acting as insulating barriers between hole-rich domains),
in this Chapter we present some novel phenomena related to
mechanical, magnetic, electric and transport properties of NGS in
underdoped single crystals. In particular, we consider chemically
induced magnetoelectric effects and flux driven temperature
oscillations of thermal expansion coefficient. We also predict a
giant enhancement of the nonlinear thermal conductivity of NGS
reaching up to $500\%$ when the intrinsically induced
chemoelectric field (created by the gradient of the chemical
potential due to segregation of hole producing oxygen vacancies)
closely matches the externally produced thermoelectric field. The
estimates of the model parameters suggest quite an optimistic
possibility to experimentally realize these promising and
important for applications effects in non-stoichiometric NGS and
artificially prepared arrays of Josephson nanocontacts}.
\vspace{5mm}

\leftline{1. INTRODUCTION} \vspace{5mm}

Inspired by new possibilities offered by the cutting-edge
nanotechnologies,  the experimental and theoretical physics of
increasingly sophisticated mesoscopic quantum devices, heavily
based on Josephson junctions (JJs) and their arrays (JJAs), is
becoming one of the most exciting and rapidly growing areas of
modern science (for reviews on charge and spin effects in
mesoscopic 2D JJs and quantum-state engineering with Josephson
devices, see, e.g., Newrock et al. 2000, Makhlin et al. 2001,
Krive et al. 2004, Sergeenkov 2006, Beloborodov et al. 2007). In
particular, a remarkable increase of the measurements technique
resolution  made it possible to experimentally detect such
interesting phenomena as flux avalanches (Altshuler and Johansen
2004) and geometric quantization (Sergeenkov and Araujo-Moreira
2004) as well as flux dominated behavior of heat capacity
(Bourgeois et al. 2005) both in JJs and JJAs.

Recently, it was realized that JJAs can be also used as quantum
channels to transfer quantum information between distant sites
(Makhlin et al. 2001, Wendin and Shumeiko 2007) through the
implementation of the so-called superconducting qubits which take
advantage of both charge and phase degrees of freedom.

Both granular superconductors and artificially prepared JJAs
proved useful in studying the numerous quantum (charging) effects
in these interesting systems, including Coulomb blockade of Cooper
pair tunneling (Iansity et al. 1988), Bloch oscillations (Haviland
et al. 1991), propagation of quantum ballistic vortices (van der
Zant 1996), spin-tunneling related effects using specially
designed $SFS$-type junctions (Ryazanov et al. 2001, Golubov et
al. 2002), novel Coulomb effects in $SINIS$-type nanoscale
junctions (Ostrovsky and Feigel'man 2004), and dynamical AC
reentrance (Araujo-Moreira et al. 1997, Barbara et al. 1999,
Araujo-Moreira et al. 2005).

At the same time, given a rather specific magnetostrictive
(Sergeenkov and Ausloos 1993) and piezomagnetic (Sergeenkov 1998b,
Sergeenkov 1999) response of Josephson systems, one can expect
some nontrivial behavior of the thermal expansion (TE) coefficient
in JJs as well (Sergeenkov et al. 2007). Of special interest are
the properties of TE in applied magnetic field. For example, some
superconductors like $Ba_{1-x}K_xBiO_3$, $BaPb_xBi_{1-x}O_3$ and
$La_{2-x}Sr_xCuO_4$ were found (Anshukova et al. 2000) to exhibit
anomalous temperature behavior of both magnetostriction and TE
which were attributed to the field-induced suppression of the
superstructural ordering in the oxygen sublattices of these
systems.

The imaging of the granular structure in underdoped
$Bi_2Sr_2CaCu_2O_{8+\delta}$ crystals (Lang et al. 2002) revealed
an apparent segregation of its electronic structure into
superconducting domains (of the order of a few nanometers) located
in an electronically distinct background. In particular, it was
found that at low levels of hole doping ($\delta <0.2$), the holes
become concentrated at certain hole-rich domains. (In this regard,
it is interesting to mention a somewhat similar phenomenon of
"chemical localization" that takes place in materials, composed of
atoms of only metallic elements, exhibiting metal-insulator
transitions, see, e.g., Gantmakher 2002.) Tunneling between such
domains leads to intrinsic nanogranular superconductivity (NGS) in
high-$T_c$ superconductors (HTS). Probably one of the first
examples of NGS was observed in $YBa_2Cu_3O_{7-\delta }$ single
crystals in the form of the so-called "fishtail" anomaly of
magnetization (Daeumling et al. 1990). The granular behavior has
been related to the 2D clusters of oxygen defects forming twin
boundaries (TBs) or dislocation walls within $CuO$ plane that
restrict supercurrent flow and allow excess flux to enter the
crystal. Indeed, there are serious arguments to consider the TB in
HTS as insulating regions of the Josephson SIS-type structure. An
average distance between boundaries is essentially less than the
grain size. In particular, the networks of localized grain
boundary dislocations with the spacing ranged from $10 nm$ to $100
nm$ have been observed (Daeumling et al. 1990) which produce
effectively continuous normal or insulating barriers at the grain
boundaries. It was also verified that the processes of the oxygen
ordering in HTS leads to the continuous change of the lattice
period along TB with the change of the oxygen content. Besides, a
destruction of bulk superconductivity in these non-stoichiometric
materials with increasing the oxygen deficiency parameter $\delta
$ was found to follow a classical percolation theory (Gantmakher
et al. 1990).

In addition to their importance for understanding the underlying
microscopic mechanisms governing HTS materials, the above
experiments can provide rather versatile tools for designing
chemically-controlled atomic scale JJs and JJAs with pre-selected
properties needed for manufacturing the modern quantum devices
(Sergeenkov 2001, Araujo-Moreira et al. 2002, Sergeenkov 2003,
Sergeenkov 2006). Moreover, as we shall see below, NGS based
phenomena can shed some light on the origin and evolution of the
so-called paramagnetic Meissner effect (PME) which manifests
itself both in high-$T_c$ and conventional superconductors (Geim
et al. 1998, De Leo and Rotoli 2002, Li 2003) and is usually
associated with the presence of $\pi$-junctions and/or
unconventional ($d$-wave) pairing symmetry.

In this Chapter we present numerous novel phenomena related to the
magnetic, electric, elastic and transport properties of Josephson
nanocontacts and NGS. The paper is organized as follows. In
Section 1, a realistic model of NGS is introduced which is based
on 2D JJAs created by a regular network of twin-boundary
dislocations with strain fields acting as an insulating barrier
between hole-rich domains (like in underdoped crystals). In
Section 2, we consider some phase-related phenomena expected to
occur in NGS, such as Josephson chemomagnetism and
magnetoconcentration effect. Section 3 is devoted to a thorough
discussion of charge-related polarization phenomena in NGS,
including such topics as chemomagnetoelectricity,
magnetocapacitance, charge analog of the "fishtail"
(magnetization) anomaly, and field-tuned weakening of the
chemically-induced Coulomb blockade. In Section 4 we present our
latest results on the influence of an intrinsic chemical pressure
(created by the gradient of the chemical potential  due to
segregation of hole producing oxygen vacancies) on temperature
behavior of the nonlinear thermal conductivity (NLTC) of NGS. In
particular, our theoretical analysis (based on the inductive model
of 2D JJAs) predicts a giant enhancement of NLTC reaching up to
$500\%$  when the intrinsically induced chemoelectric field ${\bf
E}_\mu =\frac{1}{2e}\nabla \mu$ closely matches thermoelectric
field ${\bf E}_T=S_T\nabla T$. And finally, by introducing a
concept of thermal expansion (TE) of Josephson contact (as an
elastic response of JJ to an effective stress field), in Section 5
we consider the temperature and magnetic field dependence of the
TE coefficient $\alpha (T,H)$ in a small single JJ and in a single
plaquette (a prototype of the simplest JJA). In particular, we
found that in addition to expected {\it field} oscillations due to
Fraunhofer-like dependence of the critical current, $\alpha $ of a
small single junction also exhibits strong flux driven {\it
temperature} oscillations near $T_C$. The condition under which
all the effects predicted here can be experimentally realized in
artificially prepared JJAs and NGS are also discussed. Some
important conclusions of the present study are drawn in Section 6.

\vspace{4mm} \leftline{2. MODEL OF NANOSCOPIC JOSEPHSON JUNCTION
ARRAYS } \vspace{5mm}

As is well-known, the presence of a homogeneous chemical potential
$\mu$ through a single JJ leads to the AC Josephson effect with
time dependent phase difference $\partial \phi /\partial t=\mu
/\hbar$. In this Section, we will consider some effects in
dislocation induced JJ caused by a local variation of excess hole
concentration $c({\bf x})$ under the chemical pressure (described
by inhomogeneous chemical potential $\mu ({\bf x})$) equivalent to
presence of the strain field of 2D dislocation array $\epsilon
({\bf x})$ forming this Josephson contact.

To understand how NGS manifests itself in non-stoichiometric
crystals, let us invoke an analogy with the previously discussed
dislocation models of twinning-induced superconductivity (Khaikin
and Khlyustikov 1981) and grain-boundary Josephson junctions
(Sergeenkov 1999). Recall that under plastic deformation, grain
boundaries (GBs) (which are the natural sources of weak links in
HTS), move rather rapidly via the movement of the grain boundary
dislocations (GBDs) comprising these GBs. At the same time,
observed (Daeumling et al. 1990, Lang et al. 2002, Yang et al.
1993, Moeckley et al. 1993) in HTS single crystals regular 2D
dislocation networks of oxygen depleted regions (generated by the
dissociation of $<110>$ twinning dislocations) with the size $d_0$
of a few Burgers vectors, forming a triangular lattice with a
spacing $d\ge d_0$ ranging from $10nm$ to $100nm$, can provide
quite a realistic possibility for the existence of 2D Josephson
network within $CuO$ plane. Recall furthermore that in a $d$-wave
orthorhombic $YBCO$ crystal TBs are represented by tetragonal
regions (in which all dislocations are equally spaced by $d_0$ and
have the same Burgers vector ${\bf a}$ parallel to $y$-axis within
$CuO$ plane) which produce screened strain fields (Gurevich and
Pashitskii 1997) $\epsilon ({\bf x})=\epsilon _0e^{-{\mid{{\bf
x}}\mid}/d_0}$ with ${\mid{{\bf x}}\mid}=\sqrt{x^2+y^2}$.

Though in $YBa_2Cu_3O_{7-\delta }$ the ordinary oxygen diffusion
$D=D_0e^{-U_d/k_BT}$ is extremely slow even near $T_C$ (due to a
rather high value of the activation energy $U_d$ in these
materials, typically $U_d\simeq 1eV$), in underdoped crystals
(with oxygen-induced dislocations) there is a real possibility to
facilitate oxygen transport via the so-called osmotic (pumping)
mechanism (Girifalco 1973, Sergeenkov 1995) which relates a local
value of the chemical potential (chemical pressure) $\mu ({\bf
x})=\mu (0)+\nabla \mu \cdot {\bf x}$ with a local concentration
of point defects as follows $c({\bf x})=e^{-\mu ({\bf x})/k_BT}$.
Indeed, when in such a crystal there exists a nonequilibrium
concentration of vacancies, dislocation is moved for atomic
distance $a$ by adding excess vacancies to the extraplane edge.
The produced work is simply equal to the chemical potential of
added vacancies. What is important, this mechanism allows us to
explicitly incorporate the oxygen deficiency parameter $\delta $
into our model by relating it to the excess oxygen concentration
of vacancies ${\bf c_v}\equiv c(0)$ as follows $\delta=1-{\bf
c_v}$. As a result, the chemical potential of the single vacancy
reads $\mu _v\equiv \mu (0)=-k_BT\log (1-\delta )\simeq k_BT\delta
$. Remarkably, the same osmotic mechanism was used by Gurevich and
Pashitskii (1997) to discuss the modification of oxygen vacancies
concentration in the presence of the TB strain field. In
particular, they argue that the change of $\epsilon ({\bf x})$
under an applied or chemically induced pressure results in a
significant oxygen redistribution producing a highly inhomogeneous
filamentary structure of oxygen-deficient nonsuperconducting
regions along GB (Moeckley et al. 1993) (for underdoped
superconductors, the vacancies tend to concentrate in the regions
of compressed material). Hence, assuming the following connection
between the variation of mechanical and chemical properties of
planar defects, namely $\mu ({\bf x})=K\Omega _0\epsilon ({\bf
x})$ (where $\Omega _0$ is an effective atomic volume of the
vacancy and $K$ is the bulk elastic modulus), we can study the
properties of TB induced JJs under intrinsic chemical pressure
$\nabla \mu$ (created by the variation of the oxygen doping
parameter $\delta $). More specifically, a single $SIS$ type
junction (comprising a Josephson network) is formed around TB due
to a local depression of the superconducting order parameter
$\Delta ({\bf x})\propto \epsilon({\bf x})$ over distance $d_0$
producing thus a weak link with (oxygen deficiency $\delta $
dependent) Josephson coupling $J(\delta )=\epsilon({\bf
x})J_0=J_0(\delta )e^{-{\mid{{\bf x}}\mid}/d_0}$ where $J_0(\delta
)=\epsilon _0J_0=(\mu _v/K\Omega _0 )J_0$ (here $J_0\propto \Delta
_0/R_n$ with $R_n$ being a resistance of the junction). Thus, the
present model indeed describes chemically induced NGS in
underdoped systems (with $\delta \neq 0$) because, in accordance
with the observations, for stoichiometric situation (when $\delta
\simeq 0$), the Josephson coupling $J(\delta ) \simeq 0$ and the
system loses its explicitly granular signature.

To adequately describe chemomagnetic properties
of an intrinsically granular superconductor, we employ a model of 2D
overdamped Josephson junction array which is based on the well known
Hamiltonian
\begin{equation}
{\cal H}=\sum_{ij}^NJ_{ij}(1-\cos \phi_{ij})+\sum_{ij}^N \frac{q_iq_j}{C_{ij}}
\end{equation}
and introduces a short-range interaction between
$N$ junctions (which are formed around oxygen-rich superconducting
areas with phases $\phi _i(t)$), arranged in a two-dimensional (2D)
lattice with coordinates ${\bf x_i}=(x_i,y_i)$. The areas are
separated by oxygen-poor insulating boundaries (created by TB strain
fields $\epsilon({\bf x}_{ij})$) producing a short-range Josephson
coupling $J_{ij}=J_0(\delta )e^{-{\mid{{\bf x}_{ij}}\mid}/d}$. Thus,
typically for granular superconductors, the Josephson energy of the
array varies exponentially with the distance ${\bf x}_{ij}={\bf
x}_{i}-{\bf x}_{j}$ between neighboring junctions (with $d$ being an
average junction size). As usual, the second term in the rhs of Eq.(1)
accounts for Coulomb effects where $q_i =-2en_i$ is the junction charge
with $n_i$ being the pair number operator. Naturally, the same strain
fields $\epsilon({\bf x}_{ij})$ will be responsible for dielectric properties
of oxygen-depleted regions as well via the $\delta $-dependent capacitance tensor
$C_{ij}(\delta )=C[\epsilon({\bf x}_{ij})]$.

If, in addition to the chemical pressure $\nabla \mu ({\bf
x})=K\Omega _0\nabla \epsilon ({\bf x})$, the network of
superconducting grains is under the influence of an applied
frustrating magnetic field ${\bf B}$, the total phase difference
through the contact reads
\begin{equation}
\phi _{ij}(t)=\phi ^0_{ij}+\frac{\pi w}{\Phi _0} ({\bf x}_{ij}\wedge
{\bf n}_{ij})\cdot {\bf B}+\frac{\nabla \mu \cdot {\bf
x}_{ij}t}{\hbar},
\end{equation}
where $\phi ^0_{ij}$ is the initial phase difference (see below),
${\bf n}_{ij}={\bf X}_{ij}/{\mid{{\bf X}_{ij}}\mid}$ with $ {\bf
X}_{ij}=({\bf x}_{i}+{\bf x}_{j})/2$, and $w=2\lambda _L(T)+l$
with $\lambda _L$ being the London penetration depth of
superconducting area and $l$ an insulator thickness which, within
the discussed here scenario, is simply equal to the TB thickness
(Sergeenkov 1995).

To neglect the influence of the self-field effects in a real
material, the corresponding Josephson penetration length $\lambda
_J=\sqrt{\Phi _0/2\pi \mu _0j_c w}$ must be larger than the junction
size $d$. Here $j_c$ is the critical current density of
superconducting (hole-rich) area. As we shall see below, this
condition is rather well satisfied for HTS single crystals.

Within our scenario, the sheet magnetization {\bf M} of 2D
granular superconductor is defined via the average Josephson
energy of the array
\begin{equation}
<{\cal H}>=\int_0^\tau \frac{dt}{\tau}\int \frac{d^2x}{s} {\cal
H}({\bf x},t)
\end{equation}
as follows
\begin{equation}
 {\bf M}({\bf B},\delta )\equiv -\frac{\partial
<{\cal H}>}{\partial {\bf B}},
\end{equation}
where $s=2\pi d^2$ is properly defined normalization area, $\tau$
is a characteristic Josephson time, and we made a usual
substitution $\frac{1}{N}\sum_{ij}A_{ij}(t) \to \frac{1}{s}\int
d^2x A({\bf x},t)$ valid in the long-wavelength approximation
(Sergeenkov 2002).

To capture the very essence of the superconducting analog of the
chemomagnetic effect, in what follows we assume for simplicity
that a {\it stoichiometric sample} (with $\delta \simeq 0$) does
not possess any spontaneous magnetization at zero magnetic field
(that is ${\bf M}(0,0)=0$) and that its Meissner response to a
small applied field {\bf B} is purely diamagnetic (that is ${\bf
M}({\bf B},0)\simeq -{\bf B}$). According to Eq.(4), this
condition implies $\phi _{ij}^0=2\pi m$ for the initial phase
difference with $m=0,\pm 1, \pm 2,..$.

Taking the applied magnetic field along the $c$-axis (and normal
to the $CuO$ plane), we obtain finally
\begin{equation}
{\bf M}({\bf B},\delta )=-{\bf M}_0(\delta )\frac{{\bf b}-{\bf
b}_{\mu }}{(1+{\bf b}^2)(1+({\bf b}-{\bf b}_{\mu })^2)}
\end{equation}
for the chemically-induced sheet magnetization of the 2D Josephson
network. Here ${\bf M}_0(\delta )=J_0(\delta )/{\bf B}_0$ with
$J_0(\delta )$ defined earlier, ${\bf b}={\bf B}/{\bf B}_0$, and
${\bf b}_{\mu }={\bf B}_{\mu }/{\bf B}_0\simeq (k_BT\tau /\hbar
)\delta $ where ${\bf B}_{\mu }(\delta )=(\mu _v\tau /\hbar ){\bf
B}_0$ is the chemically-induced contribution (which disappears in
optimally doped systems with $\delta \simeq 0$), and ${\bf
B}_0=\Phi _0/wd$ is a characteristic Josephson field.
\begin{SCfigure}[][ht]
 \includegraphics[width=65mm]{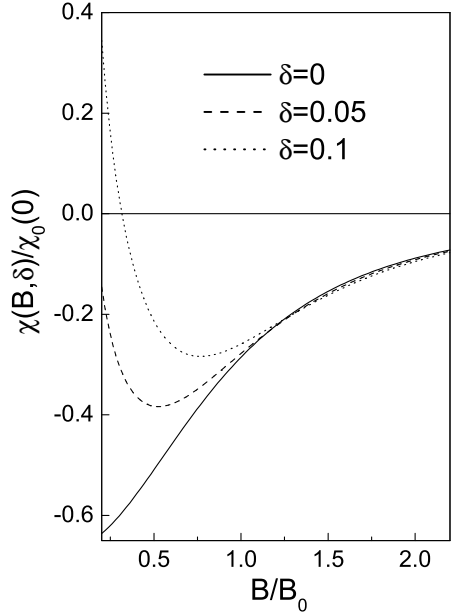}
 \caption{The susceptibility as a function of applied magnetic field
  for different values of oxygen deficiency parameter: ${\bf
\delta} \simeq 0$ (solid line), ${\bf \delta} =0.05$ (dashed
line), and ${\bf \delta} =0.1$ (dotted line). }
 \label{fig1}
 \end{SCfigure}

\begin{SCfigure}[][ht]
 \includegraphics[width=65mm]{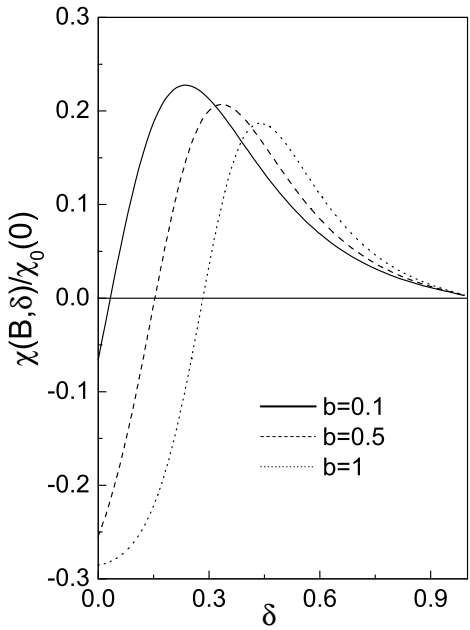}
 \caption{The oxygen deficiency induced susceptibility  for different values of
applied magnetic field (chemomagnetism). }
 \label{fig2}
 \end{SCfigure}
Fig.\ref{fig1} shows changes of the initial (stoichiometric)
diamagnetic susceptibility $\chi ({\bf B},\delta )=\partial {\bf
M}({\bf B},\delta )/\partial {\bf B}$ (solid line) with oxygen
deficiency $\delta$. As is seen, even relatively small values of
$\delta$ parameter render a low field Meissner phase strongly
paramagnetic (dotted and dashed lines). Fig.\ref{fig2} presents
concentration (deficiency) induced susceptibility $\chi ({\bf
B},\delta )/\chi _0(0)$ for different values of applied magnetic
field ${\bf b}={\bf B}/{\bf B}_0$ including a true {\it
chemomagnetic} effect (solid line). According to Eq.(5), the
initially diamagnetic Meissner effect turns paramagnetic as soon
as the chemomagnetic contribution ${\bf B}_{\mu }(\delta )$
exceeds an applied magnetic field ${\bf B}$. To see whether this
can actually happen in a real material, let us estimate a
magnitude of the chemomagnetic field ${\bf B}_{\mu }$. Typically
(Daeumling et al. 1990, Gurevich and Pashitskii 1997), for HTS
single crystals $\lambda _L(0)\approx 150nm$ and $d\simeq 10nm$,
leading to ${\bf B}_0\simeq 0.5T$. Using $\tau \simeq \hbar /\mu
_v$ and $j_c=10^{10}A/m^2$ as a pertinent characteristic time and
the typical value of the critical current density, respectively,
we arrive at the following estimate of the chemomagnetic field
${\bf B}_{\mu }(\delta )\simeq 0.5B_0$ for $\delta =0.05$. Thus,
the predicted chemically induced PME should be observable for
applied magnetic fields ${\bf B}\simeq 0.5B_0\simeq 0.25T$ which
are actually much higher than the fields needed to observe the
previously discussed piezomagnetism and stress induced PME in
high-$T_c$ ceramics (Sergeenkov 1999). Notice that for the above
set of parameters, the Josephson length $\lambda _J\simeq 1\mu m$,
which means that the assumed here small-junction approximation
(with $d\ll \lambda _J$) is valid and the so-called "self-field"
effects can be safely neglected.
\begin{SCfigure}[][ht]
 \includegraphics[width=65mm]{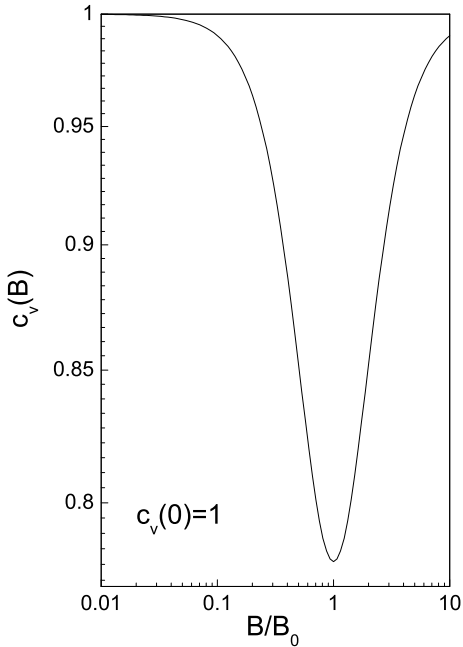}
 \caption{Magnetic field dependence of the oxygen
 vacancy concentration (magnetoconcentration effect).}
 \label{fig3}
 \end{SCfigure}
So far, we neglected a possible field dependence of the chemical
potential $\mu _v$ of oxygen vacancies. However, in high enough
applied magnetic fields ${\bf B}$, the field-induced change of the
chemical potential $\Delta \mu _v({\bf B})\equiv \mu _v({\bf
B})-\mu _v(0)$ becomes tangible and should be taken into account.
As is well-known (Abrikosov 1988, Sergeenkov and Ausloos 1999), in
a superconducting state $\Delta \mu _v({\bf B})=-{\bf M}({\bf
B}){\bf B}/n$, where ${\bf M}({\bf B})$ is the corresponding
magnetization, and $n$ is the relevant carriers number density. At
the same time, within our scenario, the chemical potential of a
single oxygen vacancy $\mu _v$ depends on the concentration of
oxygen vacancies (through deficiency parameter $\delta $). As a
result, two different effects are possible related respectively to
magnetic field dependence of $\mu _v({\bf B})$ and to its
dependence on magnetization $\mu _v({\bf M})$. The former is
nothing else but a superconducting analog of the so-called {\it
magnetoconcentration} effect which was predicted and observed in
inhomogeneously doped semiconductors (Akopyan et al.1990) with
field-induced creation of oxygen vacancies ${\bf c_v}({\bf
B})={\bf c_v}(0)\exp(-\Delta \mu _v({\bf B})/k_BT)$, while the
latter results in a "fishtail"-like behavior of the magnetization.
Let us start with the magnetoconcentration effect. Fig.\ref{fig3}
depicts the predicted field-induced creation of oxygen vacancies
${\bf c_v}({\bf B})$ using the above-obtained magnetization ${\bf
M}({\bf B},\delta )$ (see Fig.1 and Eq.(5)). We also assumed, for
simplicity, a complete stoichiometry of the system in a zero
magnetic field (with ${\bf c_v}(0)=1$). Notice that ${\bf
c_v}({\bf B})$ exhibits a maximum at ${\bf c_m}\simeq 0.23$ for
applied fields ${\bf B}={\bf B}_0$ (in agreement with the
classical percolative behavior observed in non-stoichiometric
$YBa_2Cu_3O_{7-\delta }$ samples (Daeumling et al. 1990,
Gantmakher et al. 1990, Moeckley et al. 1993). Finally, let us
show that in underdoped crystals the above-discussed osmotic
mechanism of oxygen transport is indeed much more effective than a
traditional diffusion. Using typical $YBCO$ parameters (Gurevich
and Pashitskii 1997), $\epsilon _0=0.01$, $\Omega _0=a_0^3$ with
$a_0=0.2nm$, and $K=115GPa$, we have $\mu _v(0)=\epsilon _0K\Omega
_0\simeq 1meV$ for a zero-field value of the chemical potential in
HTS crystals, which leads to creation of excess vacancies with
concentration ${\bf c_v}(0)=e^{-\mu _v(0)/k_BT}\simeq 0.75$
(equivalent to a deficiency value of $\delta (0)\simeq 0.25$) at
$T=T_C$, while the probability of oxygen diffusion in these
materials (governed by a rather high activation energy $U_d\simeq
1eV$) is extremely low under the same conditions because $D\propto
e^{-U_d/k_BT_C}\ll 1$. On the other hand, the change of the
chemical potential in applied magnetic field can reach as much as
(Sergeenkov and Ausloos 1999) $\Delta \mu _v({\bf B})\simeq
0.5meV$ for ${\bf B}=0.5T$, which is quite comparable with the
above-mentioned zero-field value of $\mu _v(0)$.
\begin{SCfigure}[][ht]
\includegraphics[width=65mm]{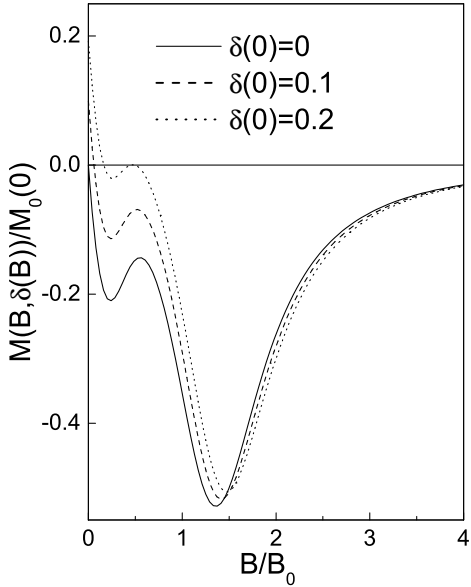}
\caption{A "fishtail"-like behavior of magnetization  in applied
magnetic field  in the presence of magnetoconcentration effect
(with field-induced oxygen vacancies ${\bf c_v}({\bf B})$, see
Fig.3) for three values of field-free deficiency parameter:
$\delta (0)\simeq 0$ (solid line), $\delta (0)=0.1$ (dashed line),
and $\delta (0)=0.2$ (dotted line).}
 \label{fig4}
 \end{SCfigure}
Let us turn now to the second effect related to the magnetization
dependence of the chemical potential $\mu _v({\bf M}({\bf B}))$.
In this case, in view of Eq.(2), the phase difference will acquire
an extra ${\bf M}({\bf B})$ dependent contribution and as a result
the r.h.s. of Eq.(5) will become a nonlinear functional of ${\bf
M}({\bf B})$. The numerical solution of this implicit equation for
the resulting magnetization $m_{f}={\bf M}({\bf B},\delta ({\bf
B}))/{\bf M}_0(0)$ is shown in Fig.\ref{fig4} for three values of
zero-field deficiency parameter $\delta (0)$. As is clearly seen,
$m_{f}$ exhibits a field-induced "fishtail"-like behavior typical
for underdoped crystals with intragrain granularity. The extra
extremum of the magnetization appears when the applied magnetic
field ${\bf B}$ matches an intrinsic chemomagnetic field ${\bf
B}_{\mu}(\delta ({\bf B}))$ (which now also depends on ${\bf B}$
via the above-discussed magnetoconcentration effect). Notice that
a "fishtail" structure of $m_{f}$ manifests itself even at zero
values of field-free deficiency parameter $\delta (0)$ (solid line
in Fig.3) thus confirming a field-induced nature of intrinsic
nanogranularity (Lang et al. 2002, Daeumling et al. 1990, Yang et
al. 1993, Gurevich and Pashitskii 1997, Moeckley et al. 1993). At
the same time, even a rather small deviation from the zero-field
stoichiometry (with $\delta (0)=0.1$) immediately brings about a
paramagnetic Meissner effect at low magnetic fields. Thus, the
present model predicts appearance of two interrelated phenomena,
Meissner paramagnetism at low fields and "fishtail" anomaly at
high fields. It would be very interesting to verify these
predictions experimentally in non-stoichiometric superconductors
with pronounced networks of planar defects.

\vspace{4mm} \leftline{3. MAGNETIC FIELD INDUCED POLARIZATION
EFFECTS IN 2D JJA} \vspace{5mm}

In this Section, within the same model of JJAs created by a
regular 2D network of twin-boundary (TB) dislocations with strain
fields acting as an insulating barrier between hole-rich domains
in underdoped crystals, we discuss charge-related effects which
are actually dual to the above-described phase-related
chemomagnetic effects. Specifically, we consider a possible
existence of a non-zero electric polarization ${\bf P}(\delta ,
{\bf B})$ (chemomagnetoelectric effect) and the related change of
the charge balance in intrinsically granular non-stoichiometric
material under the influence of an applied magnetic field. In
particular, we predict an anomalous low-field magnetic behavior of
the effective junction charge ${\bf Q}(\delta, {\bf B})$ and
concomitant magnetocapacitance ${\bf C}(\delta ,{\bf B})$ in
paramagnetic Meissner phase and a charge analog of "fishtail"-like
anomaly at high magnetic fields along with field-tuned weakening
of the chemically-induced Coulomb blockade (Sergeenkov 2007).

Recall that a conventional (zero-field) pair polarization operator
within the model under discussion reads (Sergeenkov 1997, 2002,
2007)
\begin{equation}
{\bf p}=\sum_{i=1}^Nq_i {\bf x}_{i}
\end{equation}
In view of Eqs.(1), (2) and (6), and taking into account a usual
"phase-number" commutation relation, $[\phi _i,n_j]=i\delta
_{ij}$, it can be shown that the evolution of the pair polarization operator is
determined via the equation of motion
\begin{equation}
\frac{d{\bf p}}{dt}=\frac{1}{i\hbar}\left[ {\bf p},{\cal H}\right
] =\frac{2e}{\hbar }\sum_{ij}^NJ_{ij}\sin \phi _{ij}(t){\bf x}_{ij}
\end{equation}
Resolving the above equation, we arrive at the following net value
of the magnetic-field induced longitudinal (along $x$-axis)
electric polarization ${\bf P}(\delta ,{\bf B})$ and the
corresponding effective junction charge
\begin{equation}
{\bf Q}(\delta ,{\bf B})=\frac{2eJ_0} {\hbar \tau d}
\int\limits_{0}^ {\tau }dt \int \limits_{0}^{t}dt'\int
\frac{d^2x}{S} \sin \phi ({\bf x}, t')xe^{-{\mid{{\bf x}}\mid}/d},
\end{equation}
where $S=2\pi d^2$ is properly defined normalization area, $\tau$
is a characteristic time (see below), and we made a usual
substitution $\frac{1}{N}\sum_{ij}A_{ij}(t) \to \frac{1}{S}\int
d^2x A({\bf x},t)$ valid in the long-wavelength approximation
(Sergeenkov 2002).

To capture the very essence of the superconducting analog of the
chemomagnetoelectric effect, in what follows we assume for
simplicity that a {\it stoichiometric sample} (with $\delta \simeq
0$) does not possess any spontaneous polarization at zero magnetic
field, that is ${\bf P}(0,0)=0$. According to Eq.(8), this
condition implies $\phi _{ij}^0=2\pi m$ for the initial phase
difference with $m=0,\pm 1, \pm 2,..$.

Taking the applied magnetic field along the $c$-axis (and normal
to the $CuO$ plane), we obtain finally
\begin{equation}
{\bf Q}(\delta ,{\bf B})={\bf Q}_0(\delta ) \frac{2{\tilde {\bf
b}}+{\bf b}(1-{\tilde {\bf b}}^2)}{(1+{\bf b}^2)(1+{\tilde {\bf
b}}^2)^2}
\end{equation}
for the magnetic field behavior of the effective junction charge in chemically
induced granular superconductors.
\begin{SCfigure}[][ht]
\includegraphics[width=65mm]{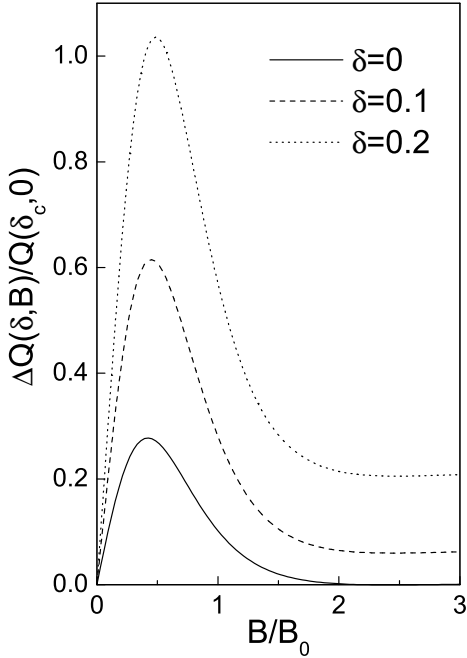}
\caption{A variation of effective junction charge with an applied
magnetic field (chemomagnetoelectric effect) for different values
of oxygen deficiency parameter: $\delta \simeq 0$ (solid line),
$\delta =0.1$ (dashed line), and $\delta=0.2$ (dotted line).}
 \label{fig5}
 \end{SCfigure}
Here ${\bf Q}_0(\delta )=e\tau J_0(\delta )/\hbar$ with
$J_0(\delta )$ defined earlier, ${\bf b}={\bf B}/{\bf B}_0$,
${\tilde {\bf b}}={\bf b}-{\bf b}_{\mu }$, and ${\bf b}_{\mu
}={\bf B}_{\mu }/{\bf B}_0\simeq (k_BT\tau /\hbar )\delta $ where
${\bf B}_{\mu }(\delta )=(\mu _v\tau /\hbar ){\bf B}_0$ is the
chemically-induced contribution (which disappears in optimally
doped systems with $\delta \simeq 0$), and ${\bf B}_0=\Phi _0/wd$
is a characteristic Josephson field.

Fig.\ref{fig5} shows changes of the initial (stoichiometric)
effective junction charge $\Delta {\bf Q}(\delta ,{\bf B})={\bf
Q}(\delta ,{\bf B})-{\bf Q}(\delta ,0)$ (solid line) with oxygen
deficiency $\delta$. According to Eq.(9), the effective charge
${\bf Q}$ changes its sign at low magnetic fields (driven by
non-zero values of $\delta$) as soon as the chemomagnetic
contribution ${\bf B}_{\mu }(\delta )$ exceeds an applied magnetic
field ${\bf B}$. This is nothing else but a charge analog of
chemically induced PME.
\begin{SCfigure}[][ht]
\includegraphics[width=65mm]{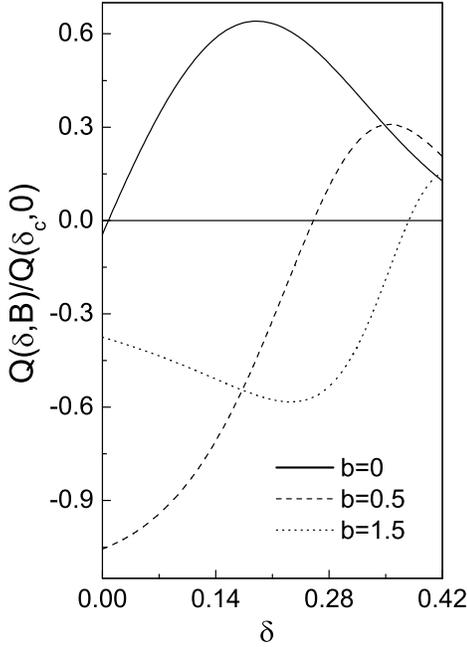}
\caption{A variation of the {\it chemomagnetoelectric} effect with
concentration (deficiency) for different values of the applied
magnetic field.}
 \label{fig6}
 \end{SCfigure}
At the same time, Fig.\ref{fig6} presents a variation of the {\it
chemomagnetoelectric} effect with concentration (deficiency) for
different values of the applied magnetic field. Notice that a
zero-field contribution (which is a true {\it chemoelectric}
effect) exhibits a maximum around $\delta _c\simeq 0.2$, in
agreement with the classical percolative behavior observed in
non-stoichiometric $YBa_2Cu_3O_{7-\delta }$ samples (Gantmakher et
al. 1990).

It is of interest also to consider the magnetic field behavior of
the concomitant effective flux capacitance ${\bf C}\equiv \tau
d{\bf Q}(\delta ,{\bf B})/d\Phi $ which in view of Eq.(9) reads
\begin{equation}
{\bf C}(\delta ,{\bf B})={\bf C}_0(\delta )\frac{1-3{\bf b}{\tilde
{\bf b}}-3{\tilde {\bf b}}^2+{\bf b}{\tilde {\bf b}}^3}{(1+{\bf
b}^2)(1+{\tilde {\bf b}}^2)^3},
\end{equation}
where $\Phi =SB$, and ${\bf C}_0(\delta )=\tau {\bf Q}_0(\delta
)/\Phi _0$.

\begin{SCfigure}[][ht]
\includegraphics[width=65mm]{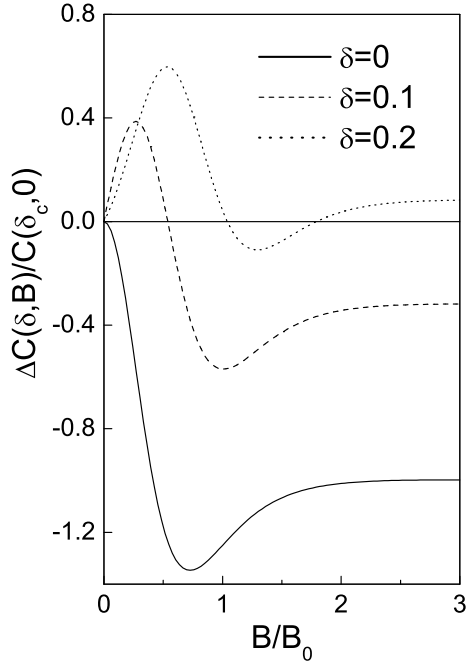}
\caption{The effective flux capacitance as a function of applied
magnetic field  for different values of oxygen deficiency
parameter: $\delta \simeq 0$ (solid line), $\delta =0.1$ (dashed
line), and $\delta=0.2$ (dotted line).}
 \label{fig7}
 \end{SCfigure}

Fig.\ref{fig7} depicts the behavior of the effective flux
capacitance $\Delta {\bf C}(\delta ,{\bf B})={\bf C}(\delta ,{\bf
B})-{\bf C}(\delta ,0)$ in applied magnetic field for different
values of oxygen deficiency parameter: $\delta \simeq 0$ (solid
line), $\delta =0.1$ (dashed line), and $\delta=0.2$ (dotted
line). Notice a decrease of magnetocapacitance amplitude and its
peak shifting with increase of $\delta$ and sign change at low
magnetic fields which is another manifestation of the charge
analog of chemically induced PME (Cf. Fig.\ref{fig5}).
Up to now, we neglected a possible field dependence of the
chemical potential $\mu _v$ of oxygen vacancies. Recall, however,
that in high enough applied magnetic fields ${\bf B}$, the
field-induced change of the chemical potential $\Delta \mu _v({\bf
B})\equiv \mu _v({\bf B})-\mu _v(0)$ becomes tangible and should
be taken into account (Abrikosov 1988, Sergeenkov and Ausloos
1999). As a result, we end up with a superconducting analog of the
so-called {\it magnetoconcentration} effect (Sergeenkov 2003) with
field induced creation of oxygen vacancies ${\bf c_v}({\bf
B})={\bf c_v}(0)\exp(-\Delta \mu _v({\bf B})/k_BT)$ which in turn
brings about a "fishtail"-like behavior of the high-field
chemomagnetization (see Section 2 for more details).
\begin{SCfigure}[][ht]
\includegraphics[width=65mm]{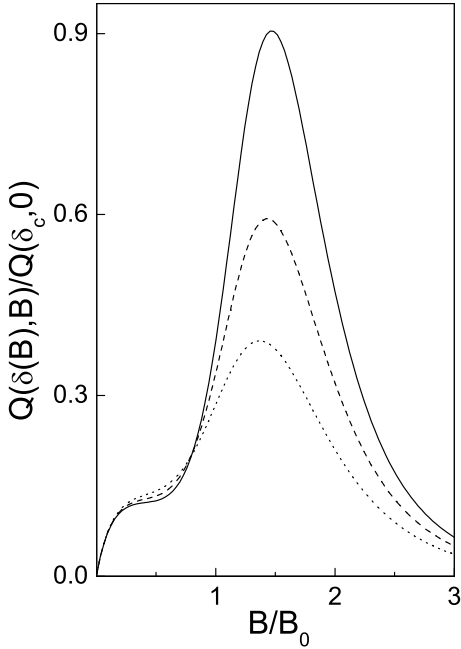}
\caption{A "fishtail"-like behavior of an effective charge in
applied magnetic field in the presence of magnetoconcentration
effect (with field-induced oxygen vacancies  $\delta ({\bf B})$)
for three values of field-free deficiency parameter (from top to
bottom):  $\delta (0)\simeq 0$ (solid line), $\delta (0)=0.1$
(dashed line), and $\delta (0)=0.2$ (dotted line).}
 \label{fig8}
 \end{SCfigure}
Fig.\ref{fig8} shows the field behavior of the effective junction
charge in the presence of the above-mentioned magnetoconcentration
effect. As it is clearly seen, ${\bf Q}(\delta ({\bf B}),{\bf B})$
exhibits a "fishtail"-like anomaly typical for previously
discussed (Sergeenkov 2003) chemomagnetization in underdoped
crystals with intragrain granularity. This more complex structure
of the effective charge appears when the applied magnetic field
${\bf B}$ matches an intrinsic chemomagnetic field ${\bf
B}_{\mu}(\delta ({\bf B}))$ (which now also depends on ${\bf B}$
via the magnetoconcentration effect). Notice that a "fishtail"
structure of ${\bf Q}(\delta ({\bf B}),{\bf B})$ manifests itself
even at zero values of field-free deficiency parameter $\delta
(0)$ (solid line in Fig.\ref{fig8}) thus confirming a
field-induced nature of intrinsic granularity.
\begin{SCfigure}[][ht]
\includegraphics[width=65mm]{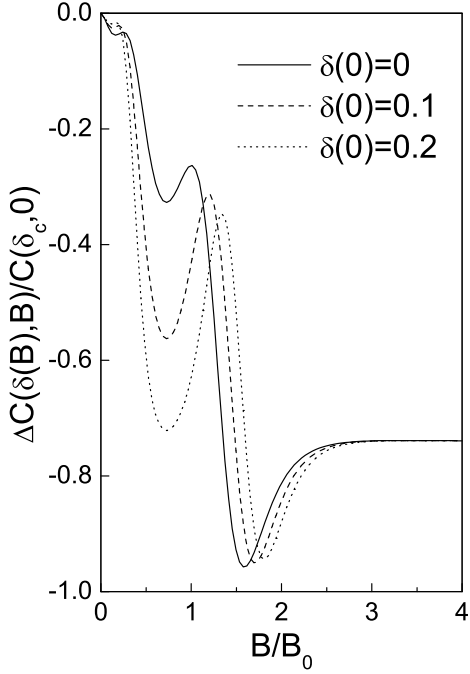}
\caption{The behavior of the effective flux capacitance
 in applied magnetic field in the presence of magnetoconcentration effect
 for three values of field-free deficiency parameter: $\delta (0)\simeq 0$ (solid line),
 $\delta (0)=0.1$ (dashed line), and $\delta (0)=0.2$ (dotted line).}
 \label{fig9}
 \end{SCfigure}
Likewise, Fig.\ref{fig9} depicts the evolution of the effective
flux capacitance $\Delta {\bf C}(\delta ({\bf B}),{\bf B})={\bf
C}(\delta ({\bf B}),{\bf B})-{\bf C}(\delta (0),0)$ in applied
magnetic field ${\bf B}$ in the presence of magnetoconcentration
effect (Cf. Fig.\ref{fig7}).

Thus, the present model predicts appearance of two interrelated
phenomena dual to the previously discussed behavior of
chemomagnetizm (see Section 2), namely a charge analog of Meissner
paramagnetism at low fields and a charge analog of "fishtail"
anomaly at high fields. To see whether these effects can be
actually observed in a real material, let us estimate an order of
magnitude of the main model parameters.

Using typical for HTS single crystals values of $\lambda _L(0)
\simeq 150nm$, $d \simeq 10nm$, and $j_c \simeq 10^{10}A/m^2$, we
arrive at the following estimates of the characteristic ${\bf B}_0
\simeq 0.5T$  and chemomagnetic ${\bf B}_{\mu }(\delta ) \simeq
0.5B_0$ fields, respectively. So, the predicted charge analog of
PME should be observable for applied magnetic fields ${\bf B} <
0.25T$. Notice that, for the above set of parameters, the
Josephson length is of the order of $\lambda _J \simeq 1\mu m$,
which means that the  small-junction approximation assumed in this
paper is valid and the "self-field" effects can be safely
neglected.

Furthermore, the characteristic frequencies $\omega \simeq \tau
^{-1}$ needed to probe the  effects suggested here are related to
the processes governed by tunneling relaxation times $\tau \simeq
\hbar /J_0(\delta )$. Since for oxygen deficiency parameter
$\delta =0.1$ the chemically-induced zero-temperature Josephson
energy in non-stoichiometric $YBCO$ single crystals is of the
order of $J_0(\delta ) \simeq k_B T_C \delta \simeq 1meV$, we
arrive at the required frequencies of $\omega \simeq 10^{13}Hz$
and at the following estimates of the effective junction charge
${\bf Q}_0 \simeq e=1.6\times 10^{-19}C$ and flux capacitance
${\bf C}_0 \simeq 10^{-18}F$. Notice that the above estimates fall
into the range of parameters used in typical experiments for
studying the single-electron tunneling effects both in JJs and
JJAs (Makhlin et al. 2001, van Bentum et al. 1988) suggesting thus
quite an optimistic possibility to observe the above-predicted
field induced effects experimentally in non-stoichiometric
superconductors with pronounced networks of planar defects or in
artificially prepared JJAs. It is worth mentioning that a somewhat
similar behavior of the magnetic field induced charge and related
flux capacitance has been observed in 2D electron systems (Chen et
al. 1994).

And finally, it can be easily verified that, in view of
Eqs.(6)-(8), the field-induced Coulomb energy of the
oxygen-depleted region within our model is given by
\begin{equation}
E_C(\delta ,{\bf B})\equiv \left < \sum_{ij}^N
\frac{q_iq_j}{2C_{ij}} \right >=\frac{{\bf Q}^2(\delta ,{\bf
B})}{2{\bf C}(\delta ,{\bf B})}
\end{equation}
with ${\bf Q}(\delta ,{\bf B})$ and ${\bf C}(\delta ,{\bf B})$
defined by Eqs. (9) and (10), respectively.

A thorough analysis of the above expression reveals that in the
PME state (when ${\bf B}\ll {\bf B}_{\mu }$) the
chemically-induced granular superconductor is in the so-called
Coulomb blockade regime (with $E_C>J_0$), while in the
''fishtail'' state (for ${\bf B}\ge {\bf B}_{\mu }$) the energy
balance tips in favor of tunneling (with $E_C<J_0$). In
particular, we obtain that $E_C(\delta ,{\bf B}=0.1{\bf B}_{\mu
})=\frac{\pi}{2}J_0(\delta )$ and $E_C(\delta ,{\bf B}={\bf
B}_{\mu })=\frac{\pi}{8}J_0(\delta )$. It would be also
interesting to check this phenomenon of field-induced weakening of
the Coulomb blockade experimentally.

\vspace{4mm} \leftline{4. GIANT ENHANCEMENT OF THERMAL
CONDUCTIVITY IN 2D JJA} \vspace{5mm}

In this Section, using a 2D model of inductive Josephson junction
arrays (created by a network of twin boundary dislocations with
strain fields acting as an insulating barrier between hole-rich
domains in underdoped crystals), we study the temperature, ${\bf
T}$, and chemical pressure, $\nabla \mu$, dependence of the
thermal conductivity (TC) $\kappa$ of an intrinsically
nanogranular superconductor. Two major effects affecting the
behavior of TC under chemical pressure are predicted: decrease of
the {\it linear} (i.e., $\nabla {\bf T}$ - independent) TC, and
giant enhancement of the {\it nonlinear} (i.e., $\nabla {\bf T}$ -
dependent) TC with $[\kappa ({\bf T},\nabla {\bf T}, \nabla \mu
)-\kappa ({\bf T},\nabla {\bf T}, 0)]/\kappa ({\bf T},\nabla {\bf
T}, 0)$ reaching $500\%$ when chemoelectric field ${\bf E}_\mu
=\frac{1}{2e}\nabla \mu $ matches thermoelectric field ${\bf
E}_T=S_T\nabla {\bf T}$. The conditions under which these effects
can be experimentally measured in non-stoichiometric high-$T_C$
superconductors are discussed.

There are several approaches for studying the thermal response of
JJs and JJAs  based on phenomenology of the Josephson effect in
the presence of thermal gradients (see, e.g., van Harlingen et al.
1980, Guttman et al. 1997, Deppe and Feldman 1994, Sergeenkov
2002, Sergeenkov 2007 and further references therein). To
adequately describe transport properties of the above-described
chemically induced nanogranular superconductor for all
temperatures and under a simultaneous influence of intrinsic
chemical pressure $\nabla \mu ({\bf x})=K\Omega _0\nabla \epsilon
({\bf x})$ and applied thermal gradient $\nabla T$, we employ a
model of 2D overdamped Josephson junction array which is based on
the following total Hamiltonian (Sergeenkov 2002)
\begin{equation}
{\cal H}(t)={\cal H}_T(t)+{\cal H}_L(t)+{\cal H}_{\mu}(t),
\end{equation}
where
\begin{equation}
{\cal H}_T(t)=\sum_{ij}^NJ_{ij}[1-\cos \phi _{ij}(t)]
\end{equation}
is the well-known tunneling Hamiltonian,
\begin{equation}
{\cal H}_L(t)=\sum_{ij}^N\frac{\Phi _{ij}^2(t)}{2L_{ij}}
\end{equation}
accounts for a mutual inductance $L_{ij}$ between grains (and
controls the normal state value of the thermal conductivity, see
below) with $\Phi _{ij}(t)=(\hbar /2e)\phi _{ij}(t)$ being the
total magnetic flux through an array, and finally
\begin{equation}
{\cal H}_{\mu }(t)=\sum_{i=1}^Nn_i(t)\delta \mu _i
\end{equation}
describes chemical potential induced contribution with $\delta \mu
_i ={\bf x}_i \nabla \mu $, and $n_i$ being the pair number
operator.

According to the above-mentioned scenario, the tunneling
Hamiltonian ${\cal H}_T(t)$ introduces a short-range
(nearest-neighbor) interaction between $N$ junctions (which are
formed around oxygen-rich superconducting areas with phases $\phi
_i(t)$), arranged in a two-dimensional (2D) lattice with
coordinates ${\bf x_i}=(x_i,y_i)$. The areas are separated by
oxygen-poor insulating boundaries (created by TB strain fields
$\epsilon({\bf x}_{ij})$) producing a short-range Josephson
coupling $J_{ij}=J_0(\delta )e^{-{\mid{{\bf x}_{ij}}\mid}/d}$.
Thus, typically for granular superconductors, the Josephson energy
of the array varies exponentially with the distance ${\bf
x}_{ij}={\bf x}_{i}-{\bf x}_{j}$ between neighboring junctions
(with $d$ being an average grain size). The temperature dependence
of chemically induced Josephson coupling is governed by the
following expression, $J_{ij}({\bf T})=J_{ij}(0)F({\bf T})$ where
\begin{equation}
F({\bf T})=\frac{\Delta ({\bf T})}{\Delta (0)}\tanh \left
[\frac{\Delta ({\bf T})}{2k_B{\bf T}}\right ]
\end{equation}
and $J_{ij}(0)=[\Delta (0)/2](R_0/R_{ij})$ with $\Delta ({\bf T})$
being the temperature dependent gap parameter, $R_0=h/4e^2$ is the
quantum resistance, and $R_{ij}$ is the resistance between grains
in their normal state.

By analogy with a constant electric field ${\bf E}$, a thermal
gradient $\nabla {\bf T}$ applied to a chemically induced JJA will
cause a time evolution of the phase difference across insulating
barriers as follows (Sergeenkov 2002)
\begin{equation}
\phi _{ij}(t)=\phi _{ij}^0+\omega _{ij}(\nabla \mu , \nabla T)t
\end{equation}
Here $\phi _{ij}^0$ is the initial phase difference (see below),
and $\omega _{ij}=2e({\bf E}_\mu -{\bf E}_T){\bf x}_{ij}/\hbar$
where ${\bf E}_\mu =\frac{1}{2e}\nabla \mu$ and ${\bf
E}_T=S_T\nabla {\bf T}$ are the induced chemoelectric and
thermoelectric fields, respectively. $S_T$ is the so-called
thermophase coefficient (Sergeenkov 1998a) which is related to the
Seebeck coefficient $S_0$ as follows, $S_T=(l/d)S_0$ (where $l$ is
a relevant sample's size responsible for the applied thermal
gradient, that is $|\nabla {\bf T}|=\Delta {\bf T}/l$).

We start our consideration by discussing the temperature behavior
of the conventional (that is {\it linear}) thermal conductivity of
a chemically induced nanogranular superconductor paying a special
attention to its evolution with a mutual inductance $L_{ij}$. For
simplicity, in what follows we limit our consideration to the
longitudinal component of the total thermal flux ${\bf Q}(t)$
which is defined (in a q-space representation) via the total
energy conservation law as follows
\begin{equation}
{\bf Q}(t)\equiv \lim_{{\bf q} \to 0} \left [i\frac{{\bf q}}{{\bf
q}^2}{\dot{\cal H}_{\bf q}}(t)\right ],
\end{equation}
where ${\dot{\cal H}_{\bf q}}=\partial {\cal H}_{\bf q}/\partial
t$ with
\begin{equation}
{\cal H}_{\bf q}(t)=\frac{1}{s}\int d^2x e^{i{\bf q}{\bf x}}{\cal
H}({\bf x},t)
\end{equation}
Here $s=2\pi d^2$ is properly defined normalization area, and we
made a usual substitution $\frac{1}{N}\sum_{ij}A_{ij}(t) \to
\frac{1}{s}\int d^2x A({\bf x},t)$ valid in the long-wavelength
approximation (${\bf q} \to 0$).

In turn, the heat flux ${\bf Q}(t)$ is related to the {\it linear}
thermal conductivity (LTC) tensor $\kappa _{\alpha \beta}$ by the
Fourier law as follows (hereafter, $\{\alpha ,\beta \}=x,y,z$)
\begin{equation}
\kappa _{\alpha \beta}({\bf T},\nabla \mu )\equiv
-\frac{1}{V}\left [\frac{\partial \overline{<{\bf
Q}_{\alpha}>}}{\partial (\nabla _{\beta}{\bf T})}\right ]_{\nabla
{\bf T}=0},
\end{equation}
where
\begin{equation}
\overline{<{\bf Q}_{\alpha}>}=\frac{1}{\tau}\int_0^\tau dt<{\bf
Q}_{\alpha}(t)>
\end{equation}
Here $V$ is sample's volume, $\tau$ is a characteristic Josephson
tunneling time for the network, and $<...>$ denotes the
thermodynamic averaging over the initial phase differences $\phi
_{ij}^0$
\begin{equation}
<A(\phi _{ij}^0)>=\frac{1}{Z}\int_0^{2\pi}\prod _{ij} d\phi
_{ij}^0 A(\phi _{ij}^0)e^{-\beta H_0}
\end{equation}
with an effective Hamiltonian
\begin{equation}
H_0[\phi _{ij}^0]=\int_0^{\tau}\frac{dt}{\tau}\int
\frac{d^2x}{s}{\cal H}({\bf x},t)
\end{equation}
Here, $\beta =1/k_B{\bf T}$, and $Z=\int_0^{2\pi}\prod _{ij}d\phi
_{ij}^0 e^{-\beta H_0}$ is the partition function. The
above-defined averaging procedure allows us to study the
temperature evolution of the system.

Taking into account that in JJAs (Eichenberger et al. 1996)
$L_{ij}\propto R_{ij}$, we obtain $L_{ij}=L_{0}\exp({\mid{{\bf
x}_{ij}}\mid}/d)$ for the explicit $x$-dependence of the weak-link
inductance in our model. Finally, in view of Eqs.(12)-(23), and
making use of the usual "phase-number" commutation relation,
$[\phi _i,n_j]=i\delta _{ij}$, we find the following analytical
expression for the temperature and chemical gradient dependence of
the electronic contribution to {\it linear} thermal conductivity
of a granular superconductor
\begin{equation}
\kappa _{\alpha \beta}({\bf T},\nabla \mu )=\kappa _0[\delta
_{\alpha \beta}\eta ({\bf T},\epsilon )+\beta _L({\bf T})\nu ({\bf
T},\epsilon )f_{\alpha \beta}(\epsilon )]
\end{equation}
where
\begin{equation}
f_{\alpha \beta}(\epsilon )=\frac{1}{4}\left [\delta _{\alpha
\beta}A(\epsilon)-\epsilon _{\alpha}\epsilon
_{\beta}B(\epsilon)\right ]
\end{equation}
with
\begin{equation}
A(\epsilon)=\frac{5+3\epsilon ^2}{(1+\epsilon
^2)^2}+\frac{3}{\epsilon}\tan ^{-1}\epsilon
\end{equation}
and
\begin{equation}
B(\epsilon)=\frac{3\epsilon ^4+8\epsilon ^2-3}{\epsilon
^2(1+\epsilon ^2)^3}+\frac{3}{\epsilon ^3}\tan ^{-1}\epsilon
\end{equation}
Here, $\kappa _0=Nd^2S_T\Phi _0/VL_{0}$, $\beta _L({\bf T})=2\pi
I_C({\bf T}) L_{0}/\Phi _0$ with $I_C({\bf T})=(2e/\hbar )J({\bf
T})$ being the critical current; $\epsilon \equiv \sqrt{\epsilon
_x^2+\epsilon _y^2+\epsilon _z^2}$ with $\epsilon _{\alpha} ={\bf
E}_{\mu} ^{\alpha} /{\bf E}_0$ where ${\bf E}_0=\hbar /2ed\tau$ is
a characteristic field. In turn, the above-introduced "order
parameters" of the system, $\eta ({\bf T},\epsilon )\equiv <\phi
_{ij}^0>$ and $\nu ({\bf T},\epsilon )\equiv <\sin \phi _{ij}^0>$,
are defined as follows
\begin{equation}
\eta ({\bf T}, \epsilon
)=\frac{\pi}{2}-\frac{4}{\pi}\sum_{n=0}^{\infty}\frac{1}{(2n+1)^2}
\left [\frac{I_{2n+1}(\beta _\mu )}{I_0(\beta _\mu )}\right ]
\end{equation}
and
\begin{equation}
\nu ({\bf T}, \epsilon )= \frac{\sinh \beta _\mu }{\beta _\mu
I_0(\beta _\mu )},
\end{equation}
where
\begin{equation}
\beta _\mu ({\bf T},\epsilon )=\frac{\beta J({\bf T})}{2}\left
(\frac{1}{1+\epsilon ^2}+\frac{1}{\epsilon}\tan ^{-1}\epsilon
\right)
\end{equation}
Here $J({\bf T})$ is given by Eq.(17), and $I_n(x)$ stand for the
modified Bessel functions.

Turning to the discussion of the obtained results, we start with a
more simple zero-pressure case. The relevant parameters affecting
the behavior of the LTC in this particular case include the mutual
inductance $L_{0}$ and the normal state resistance between grains
$R_n$. For the temperature dependence of the Josephson energy (see
Eq.(17)), we used the well-known (Sergeenkov 2002) approximation
for the BCS gap parameter, valid for all temperatures, $\Delta
({\bf T})=\Delta (0)\tanh \left (\gamma \sqrt{\frac{{\bf T_C}-{\bf
T}}{{\bf T}}}\right )$ with $\gamma =2.2$.
\begin{figure}[ht]
\centerline{\includegraphics[width=60mm]{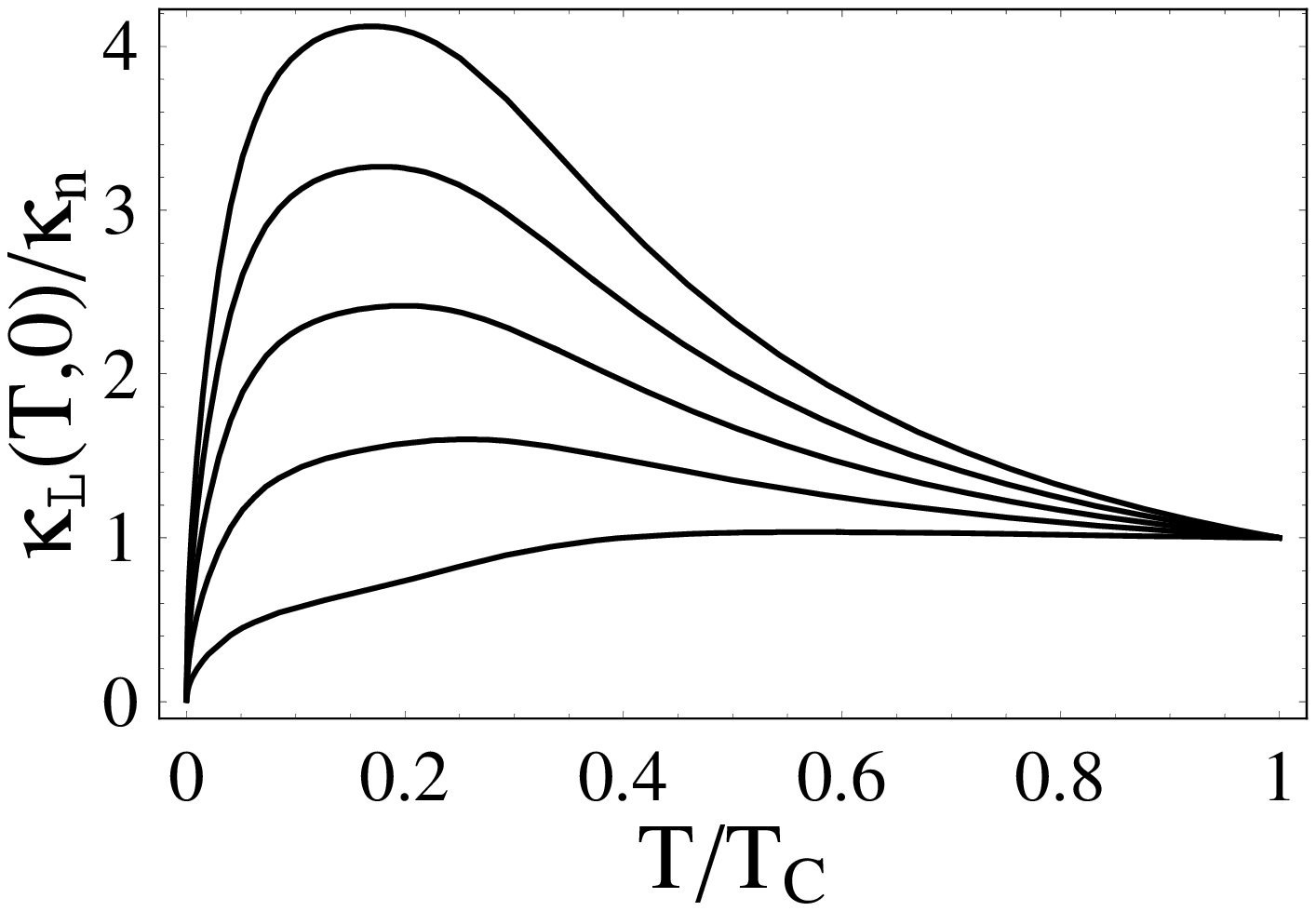} \hspace{10mm}
\includegraphics[width=60mm]{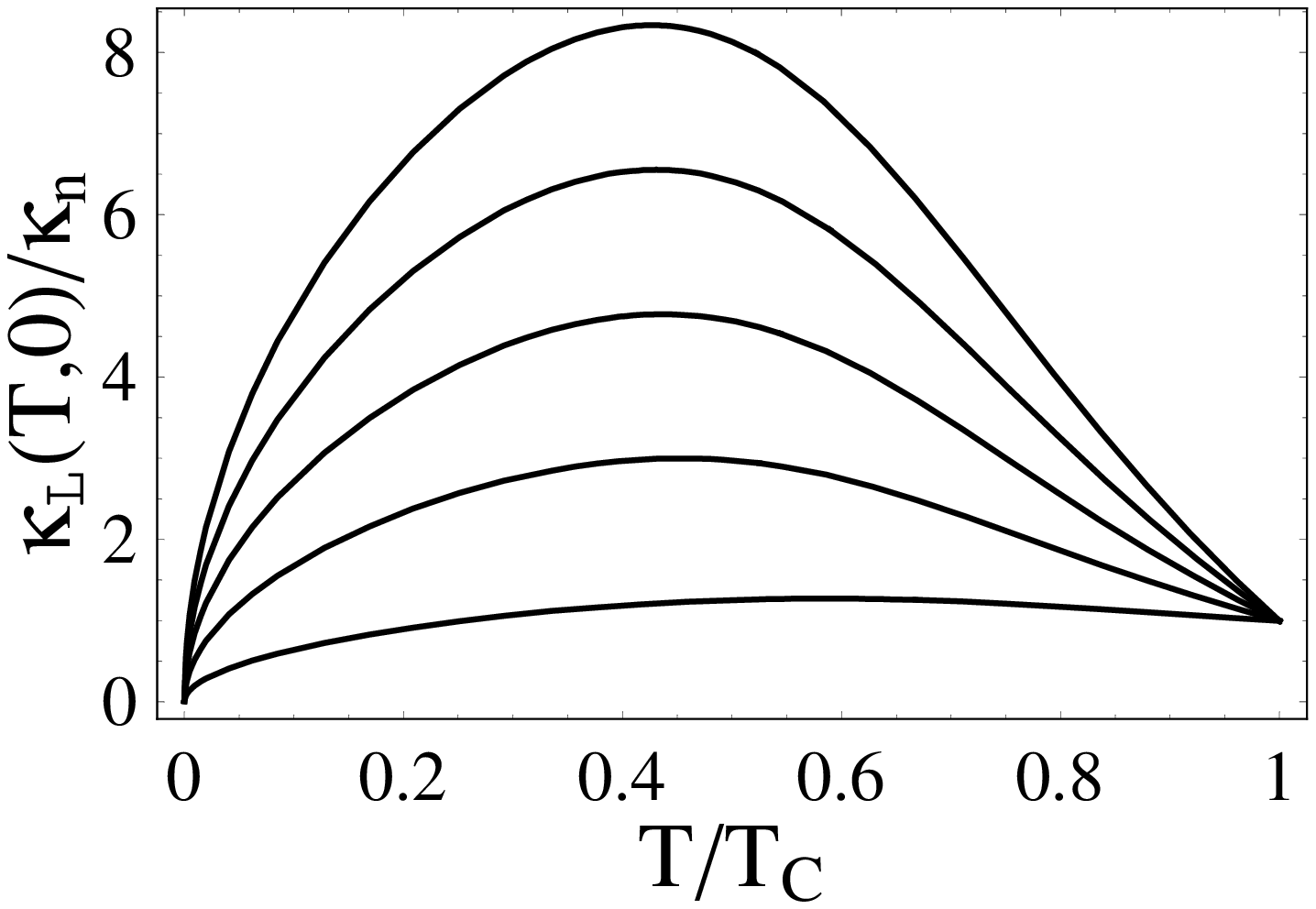}}
\caption{Temperature dependence of the zero-pressure ($\nabla \mu
=0$) {\it linear} thermal conductivity for $r_n=0.1$ (left) and
$r_n=1$ (right) for different values of the SQUID parameter (from
bottom to top): $\beta _L(0)=1,3,5,7$, and $9$.}
 \label{fig10}
 \end{figure}
Despite a rather simplified nature of our model, it seems to quite
reasonably describe the behavior of the LTC for all temperatures.
Indeed, in the absence of intrinsic chemical pressure ($\nabla
\mu=0$), the LTC is isotropic (as expected), $\kappa _{\alpha
\beta}({\bf T},0)=\delta _{\alpha \beta}\kappa _L({\bf T},0)$
where $\kappa _L({\bf T},0)=\kappa _0[\eta ({\bf T},0)+2\beta
_L({\bf T})\nu ({\bf T},0)]$ vanishes at zero temperature and
reaches a normal state value $\kappa _n\equiv \kappa _L({\bf
T_C},0)=(\pi /2)\kappa _0$ at ${\bf T}={\bf T_C}$. Fig.\ref{fig10}
shows the temperature dependence of the normalized LTC $\kappa _L
({\bf T},0)/\kappa _n$ for different values of the so-called SQUID
parameter $\beta _L(0)=2\pi I_C(0)L_{0}/\Phi _0$ (increasing from
the bottom to the top) and for two values of the resistance ratio
$r_n=R_0/R_n=0.1$ and $r_n=R_0/R_n=1$. First of all, with
increasing of the SQUID parameter, the LTC evolves from a
flat-like pattern (for a relatively small values of $L_{0}$) to a
low-temperature maximum (for higher values of $\beta _L(0)$).
Notice that the peak temperature ${\bf T}_p$ is practically
insensitive to the variation of inductance parameter $L_{0}$ while
being at the same time strongly influenced by resistivity $R_n$.
Indeed, as it is clearly seen in Fig.\ref{fig10},  a different
choice of $r_n$ leads to quite a tangible shifting of the maximum.
Namely, the smaller is the normal resistance between grains $R_n$
(or the better is the quality of the sample) the higher is the
temperature at which the peak is developed. As a matter of fact,
the peak temperature ${\bf T}_p$ is related to the so-called
phase-locking temperature ${\bf T}_J$ (which marks the
establishment of phase coherence between the adjacent grains in
the array and always lies below a single grain superconducting
temperature ${\bf T_C}$) which is usually defined via an average
(per grain) Josephson coupling energy as $J({\bf T}_J,r_n)=k_B{\bf
T}_J$. Indeed, it can be shown analytically that for ${\bf
T}_J<{\bf T}<{\bf T_C}$, ${\bf T}_J(r_n) \simeq r_n{\bf T_C}$.

\begin{SCfigure}[][ht]
\includegraphics[width=67mm]{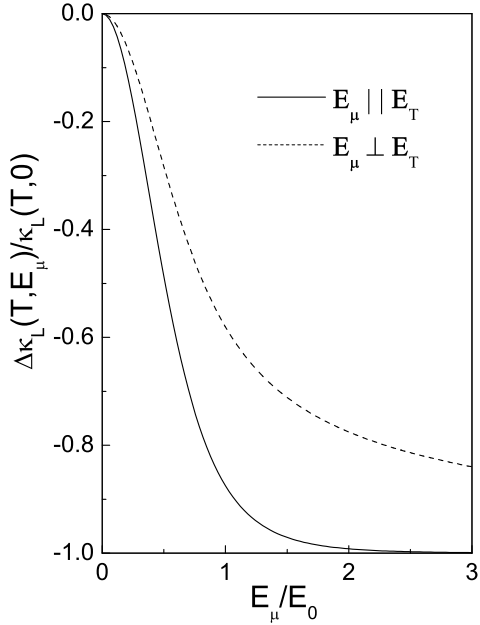}
\caption{The dependence of the {\it linear} thermal conductivity
on the chemical pressure for parallel (${\bf \nabla \mu} \| \nabla
{\bf T}$) and perpendicular (${\bf \nabla \mu} \bot \nabla {\bf
T}$) configurations.}
 \label{fig11}
 \end{SCfigure}

Turning to the discussion of the LTC behavior under chemical
pressure, let us assume, for simplicity, that $\nabla \mu =(\nabla
_x\mu ,0,0)$ with oxygen-deficiency parameter $\delta$ controlled
chemical pressure $\nabla _x\mu \simeq \mu _v(\delta )/d$, and
$\nabla {\bf T}=(\nabla _x{\bf T}, \nabla _y{\bf T},0)$. Such a
choice of the external fields allows us to consider both parallel
$\kappa _{xx}({\bf T},\nabla \mu)$ and perpendicular $\kappa
_{yy}({\bf T},\nabla \mu)$ components of the LTC corresponding to
the two most interesting configurations, ${\bf \nabla \mu} \|
\nabla {\bf T}$ and ${\bf \nabla \mu} \bot \nabla {\bf T}$,
respectively. Fig.\ref{fig11} demonstrates the predicted chemical
pressure dependence of the normalized LTC $\Delta \kappa _L({\bf
T},\nabla \mu)=\kappa _L({\bf T},\nabla \mu)-\kappa _L({\bf T},
0)$ for both configurations taken at ${\bf T}=0.2{\bf T_C}$ (with
$r_n=0.1$ and $\beta _L(0)=1$). First of all, we note that both
components of the LTC are {\it decreasing} with increasing of the
pressure ${\bf E}_\mu/{\bf E}_0=\mu _v(\delta )\tau /\hbar$. And
secondly, the normal component $\kappa _{yy}$ decreases more
slowly than the parallel one $\kappa _{xx}$, suggesting thus some
kind of anisotropy in the system. In view of the structure of
Eq.(25), the same behavior is also expected for the temperature
dependence of the chemically-induced LTC, that is $\Delta \kappa
_L({\bf T},\nabla \mu)/\kappa _L({\bf T},0)<0$ for all gradients
and temperatures. In terms of the absolute values, for ${\bf
T}=0.2{\bf T_C}$ and ${\bf E}_\mu={\bf E}_0$, we obtain $[\Delta
\kappa _L({\bf T},\nabla \mu)/\kappa _L({\bf T},0)]_{xx}=90\%$ and
$[\Delta \kappa _L({\bf T},\nabla \mu)/\kappa _L({\bf
T},0)]_{yy}=60\%$ for {\it attenuation} of LTC under chemical
pressure.
Let us turn now to the most intriguing part of this Section and
consider a {\it nonlinear} generalization of the Fourier law and
very unusual behavior of the resulting {\it nonlinear} thermal
conductivity (NLTC) under the influence of chemical pressure. In
what follows, by the NLTC we understand a $\nabla {\bf
T}$-dependent thermal conductivity $\kappa _{\alpha
\beta}^{NL}({\bf T},{\bf \nabla \mu})\equiv \kappa _{\alpha \beta
}({\bf T},{\bf \nabla \mu};\nabla {\bf T})$ which is defined as
follows
\begin{equation}
\kappa _{\alpha \beta}^{NL}({\bf T},{\bf \nabla \mu})\equiv
-\frac{1}{V}\left [\frac{\partial \overline{<{\bf
Q}_{\alpha}>}}{\partial (\nabla _{\beta}{\bf T})}\right ]_{\nabla
{\bf T}\neq 0}
\end{equation}
with $\overline{<{\bf Q}_{\alpha}>}$ given by Eq.(21).

Repeating the same procedure as before, we obtain finally for the
relevant components of the NLTC tensor
\begin{equation}
\kappa _{\alpha \beta}^{NL}({\bf T},{\bf \nabla \mu})=\kappa
_0[\delta _{\alpha \beta}\eta ({\bf T},\epsilon _{eff})+\beta
_L({\bf T})\nu ({\bf T},\epsilon _{eff})D_{\alpha \beta}(\epsilon
_{eff})],
\end{equation}
where
\begin{equation}
D_{\alpha \beta}(\epsilon _{eff})=f_{\alpha \beta}(\epsilon
_{eff})+ \epsilon _{T}^{\gamma}g_{\alpha \beta \gamma} (\epsilon
_{eff})
\end{equation}
with
\begin{equation}
g_{\alpha \beta \gamma} (\epsilon )=\frac{1}{8}[(\delta _{\alpha
\beta}\epsilon _{\gamma}+ \delta _{\alpha \gamma}\epsilon _{\beta}
+ \delta _{\gamma \beta}\epsilon _{\alpha})B(\epsilon )
 + 3\epsilon _{\alpha} \epsilon _{\beta} \epsilon _{\gamma} C(\epsilon )]
\end{equation}
and
\begin{equation}
C(\epsilon)=\frac{3+11\epsilon ^2-11\epsilon ^4-3\epsilon
^6}{\epsilon ^4(1+\epsilon ^2)^4}-\frac{3}{\epsilon ^5}\tan
^{-1}\epsilon
\end{equation}
Here, $\epsilon _{eff}^{\alpha}=\epsilon _{\mu}^{\alpha} -\epsilon
_T^{\alpha}$ where $\epsilon _{\mu}^{\alpha}={\bf E}_\mu
^{\alpha}/{\bf E}_0$ and $\epsilon _T^{\alpha}={\bf
E}_T^{\alpha}/{\bf E}_0$ with ${\bf E}_T^{\alpha}=S_T\nabla
_{\alpha}{\bf T}$; other parameters ($\eta$, $\nu$, $B$ and
$f_{\alpha \beta}$) are the same as before but with $\epsilon \to
\epsilon _{eff}$.
\begin{SCfigure}[][ht]
\includegraphics[width=67mm]{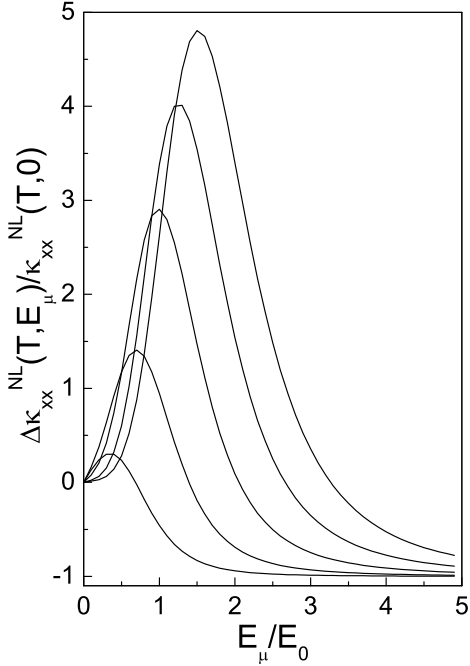}
\caption{The dependence of the {\it nonlinear} thermal
conductivity on the chemical pressure for different values of the
applied thermal gradient $\epsilon _T=S_T\nabla {\bf T}/{\bf E}_0$
($\epsilon _T=0.2, 0.4, 0.6, 0.8$, and $1.0$, increasing from
bottom to top).}
 \label{fig12}
 \end{SCfigure}
As expected, in the limit ${\bf E}_T \to 0$ (or when ${\bf E}_\mu
\gg {\bf E}_T$), from Eq.(32) we recover all the results obtained
in the previous section for the LTC. Let us see now what happens
when thermoelectric field ${\bf E}_T=S_T\nabla {\bf T}$ becomes
comparable with chemoelectric field ${\bf E_\mu}$. Fig.\ref{fig12}
depicts the resulting chemical pressure dependence of the parallel
component of the NLTC tensor $\Delta \kappa _{xx}^{NL}({\bf
T},{\bf E}_\mu)=\kappa _{xx}^{NL}({\bf T},{\bf E}_\mu)-\kappa
_{xx}^{NL}({\bf T}, 0)$ for different values of the dimensionless
parameter $\epsilon _T={\bf E}_T/{\bf E}_0$ (the other parameters
are the same as before). As is clearly seen from this picture, in
a sharp contrast with the pressure behavior of the previously
considered LTC, its {\it nonlinear} analog evolves with the
chemoelectric field quite differently. Namely, NLTC strongly {\it
increases} for small pressure values (with ${\bf E}_\mu <{\bf
E}_m$), reaches a pronounced maximum at ${\bf E}_\mu ={\bf
E}_m=\frac{3}{2}{\bf E}_T$, and eventually declines at higher
values of $_\mu$ (with ${\bf E}_\mu >{\bf E}_m$). Furthermore, as
it directly follows from the very structure of Eq.(32), a similar
"reentrant-like" behavior of the {\it nonlinear} thermal
conductivity is expected for its temperature dependence as well.
Even more remarkable is the absolute value of the pressure-induced
enhancement. According to Fig.\ref{fig12}, it is easy to estimate
that near maximum (with ${\bf E}_\mu ={\bf E}_m$ and ${\bf
E}_T={\bf E}_0$) one gets $\Delta \kappa _{xx}^{NL}({\bf T},{\bf
E}_\mu)/\kappa _{xx}^{NL}({\bf T},0)\simeq 500\%$.

To understand the above-obtained rather unusual results, let us
take a closer look at the chemoelectric field induced behavior of
the Josephson voltage in our system (see Eq.(17)). Clearly, strong
heat conduction requires establishment of a quasi-stationary (that
is nearly zero-voltage) regime within the array. In other words,
the maximum of the thermal conductivity under chemical pressure
should correlate with a minimum of the total voltage in the
system, $V(\nabla \mu)\equiv (\frac{\hbar}{2e})<\frac{\partial
\phi _{ij}(t)}{\partial t}>=V_0(\epsilon -\epsilon _T)$ where
$\epsilon \equiv {\bf E}_\mu/{\bf E}_0$ and $V_0={\bf E}_0d=\hbar
/2e\tau $ is a characteristic voltage. For linear TC (which is
valid only for small thermal gradients with $\epsilon _T\equiv
{\bf E}_T/{\bf E}_0\ll 1 $), the average voltage through an array
$V_L(\nabla \mu)\simeq V_0({\bf E}_\mu/{\bf E}_0)$ has a minimum
at zero chemoelectric field (where LTC indeed has its maximum
value, see Fig.11) while for nonlinear TC (with $\epsilon _T\simeq
1$) we have to consider the total voltage $V(\nabla \mu)$ which
becomes minimal at ${\bf E}_\mu={\bf E}_T$ (in a good agreement
with the predictions for NLTC maximum which appears at ${\bf
E}_\mu=\frac{3}{2}{\bf E}_T$, see Fig.\ref{fig12}).

To complete our study, let us estimate an order of magnitude of
the main model parameters. Starting with chemoelectric fields
${\bf E}_\mu$ needed to observe the above-predicted nonlinear
field effects in nanogranular superconductors, we notice that
according to Fig.\ref{fig12}, the most interesting behavior of
NLTC takes place for ${\bf E}_\mu \simeq {\bf E}_0$. Using typical
$YBCO$ parameters, $\epsilon _0=0.01$, $\Omega _0=a_0^3$ with
$a_0=0.2nm$, and $K=115GPa$, we have $\mu _v=\epsilon _0K\Omega
_0\simeq 1meV$ for an estimate of the chemical potential in HTS
crystals, which defines the characteristic Josephson tunneling
time $\tau \simeq \hbar /\mu _v \simeq 5\times 10^{-11}s$ and, at
the same time, leads to creation of excess vacancies with
concentration $c_v=e^{-\mu _v/k_B{\bf T}}\simeq 0.75$ at ${\bf
T}=0.2{\bf T_C}$ (equivalent to a deficiency value of $\delta
\simeq 0.25$). Notice that in comparison with this linear defects
mediated channeling (osmotic) mechanism, the probability of the
conventional oxygen diffusion in these materials $D\propto
e^{-U_d/k_BT}$ (governed by a rather high activation energy
$U_d\simeq 1eV$) is extremely low under the same conditions ($D\ll
1$).

Furthermore, taking $d\simeq 10nm$ for typical values of the
average "grain" size (created by oxygen-rich superconducting
regions), we get ${\bf E}_0=\hbar /2ed\tau \simeq 5\times
10^{5}V/m$ and $|\nabla \mu |=\mu _v/d\simeq 10^{6}eV/m$ for the
estimates of the characteristic field and chemical potential
gradient (intrinsic chemical pressure), respectively. On the other
hand, the maximum of NLTC occurs when this field nearly perfectly
matches an "intrinsic" thermoelectric field ${\bf E}_T=S_T\nabla
{\bf T}$ induced by an applied thermal gradient, that is when
${\bf E}_\mu \simeq {\bf E}_0\simeq {\bf E}_T$. Recalling that
$S_T=(l/d)S_0$ and using $S_0 \simeq 0.5\mu V/K$ and $l \simeq
0.5mm$ for an estimate of the {\it linear} Seebeck coefficient and
a typical sample's size, we obtain $\nabla {\bf T}\simeq {\bf
E}_0/S_T\simeq 2\times 10^6K/m$ for the characteristic value of
applied thermal gradient needed to observe the predicted here
giant chemical pressure induced effects. Let us estimate now the
absolute value of the linear thermal conductivity governed by the
intrinsic Josephson junctions. Recall that within our model the
scattering of normal electrons is due to the presence of mutual
inductance between the adjacent grains $L_0$ which is of the order
of $L_0\simeq \mu _0d\simeq 1fH$ assuming $d=10nm$ for an average
"grain" size. In the absence of chemical pressure effects, the
temperature evolution of LTC is given by $\kappa _L({\bf
T},0)=\kappa _0[\eta ({\bf T},0)+2\beta _L({\bf T})\nu ({\bf
T},0)]$ where $\kappa _0=Nd^2S_T\Phi _0/VL_{0}$. Assuming $V\simeq
Nd^2l$ for the sample's volume, using the above-mentioned
expression for $S_T$, and taking $\beta _L(0)=5$ and $r_n=0.1$ for
the value of the SQUID parameter and the resistance ratio, we
obtain $\kappa _L(0.2{\bf T_C},0)\simeq 1W/mK$ for an estimate of
the maximum of the LTC (see Fig.\ref{fig10}).

And finally, it is worth comparing the above estimates for
inductively coupled grains (Sergeenkov 2002) with the estimates
for capacitively coupled grains (Sergeenkov 2007) where the
scattering of normal electrons is governed by the Stewart-McCumber
parameter $\beta _C({\bf T})=2\pi I_C({\bf T})C_{0}R_n^2/\Phi _0$
due to the presence of the normal resistance $R_n$ and mutual
capacitance $C_0$ between the adjacent grains. The latter is
estimated to be $C_0\simeq 1aF$ using $d=10nm$ for an average
"grain" size. Furthermore, the critical current $I_C(0)$ can be
estimated via the critical temperature ${\bf T_C}$ as follows,
$I_C(0)\simeq 2\pi k_B{\bf T_C}/\Phi _0$ which gives $I_C(0)\simeq
10\mu A$ (for ${\bf T_C}\simeq 90K$) and leads to $\beta
_C(0)\simeq 3$ for the value of the Stewart-McCumber parameter
assuming $R_n\simeq R_0$ for the normal resistance which, in turn,
results in $q\simeq \Phi _0/R_n\simeq 10^{-19}C$ and
$E_C=q^2/2C_0\simeq 0.1eV$ for the estimates of the "grain" charge
and the Coulomb energy. Using the above-mentioned expressions for
$S_0$ and $\beta _C(0)$, we obtain $\kappa _L \simeq 10^{-3}W/mK$
for the maximum of the capacitance controlled LTC which is
actually much smaller than a similar estimate obtained above for
inductance controlled $\kappa _L$ (Sergeenkov 2002) but at the
same time much higher than phonon dominated heat transport in
granular systems (Deppe and Feldman 1994).

\vspace{4mm} \leftline{5. THERMAL EXPANSION OF A SINGLE JOSEPHSON
CONTACT AND 2D JJA} \vspace{5mm}

In this Section, by introducing a concept of thermal expansion
(TE) of a Josephson junction as an elastic response to an
effective stress field, we study (both analytically and
numerically) the temperature and magnetic field dependence of TE
coefficient $\alpha $ in a single small junction and in a square
array. In particular, we found (Sergeenkov et al. 2007) that in
addition to {\it field} oscillations due to Fraunhofer-like
dependence of the critical current, $\alpha $ of a small single
junction also exhibits strong flux driven {\it temperature}
oscillations near ${\bf T_C}$. We also numerically simulated
stress induced response of a closed loop with finite
self-inductance (a prototype of an array) and found that $\alpha $
of a $5\times 5$ array may still exhibit temperature oscillations
if the applied magnetic field ${\bf H}$ is strong enough to
compensate for the screening induced effects.

Since thermal expansion coefficient $\alpha ({\bf T},{\bf H})$ is
usually measured using mechanical dilatometers (Nagel et al.
2000), it is natural to introduce TE as an elastic response of the
Josephson contact to an effective stress field $\sigma $
(D'yachenko et al. 1995, Sergeenkov 1998b, Sergeenkov 1999).
Namely, we define the TE coefficient (TEC) $\alpha ({\bf T},{\bf
H})$ as follows:
\begin{equation}
\alpha ({\bf T},{\bf H})=\frac{d \epsilon }{d{\bf T}}
\label{alphadefinition}
\end{equation}
where an appropriate strain field $\epsilon$ in the contact area
is related to the Josephson energy $E_J$ as follows ($V$ is the
volume of the sample):
\begin{equation}
\epsilon =-\frac{1}{V}\left [\frac{dE_J}{d\sigma}\right ]_{\sigma
=0}
\end{equation}
For simplicity and to avoid self-field effects, we start with a
small Josephson contact of length $w<\lambda _J$ ($\lambda
_J=\sqrt{\Phi _0/\mu _0dj_{c}}$ is the Josephson penetration
depth) placed in a strong enough magnetic field (which is applied
normally to the contact area) such that ${\bf H}>\Phi _0/2\pi
\lambda _Jd$, where $d=2\lambda _{L}+t$, $\lambda _{L}$ is the
London penetration depth, and $t$ is an insulator thickness.

The Josephson energy of such a contact in applied magnetic field
is governed  by a Fraunhofer-like dependence of the critical
current (Orlando and Delin 1991):
\begin{equation}
E_J=J\left (1-\frac{\sin \varphi}{\varphi}\cos \varphi _0\right ),
 \label{Josenergy}
\end{equation}
where $\varphi =\pi \Phi /\Phi _0$ is the frustration parameter
with $\Phi ={\bf H}wd$ being the flux through the contact area,
$\varphi _0$ is the initial phase difference through the contact,
and $J\propto e^{-t/\xi}$ is the zero-field tunneling Josephson
energy with $\xi$ being a characteristic (decaying) length and $t$
the thickness of the insulating layer. The self-field effects
(screening), neglected here, will be considered later for an
array.

Notice that in non-zero applied magnetic field ${\bf H}$, there
are two stress-induced contributions to the Josephson energy
$E_J$, both related to decreasing of the insulator thickness under
pressure. Indeed, according to the experimental data (D'yachenko
et al. 1995), the tunneling dominated critical current $I_C$ in
granular high-$T_C$ superconductors was found to exponentially
increase under compressive stress, viz. $I_C(\sigma
)=I_C(0)e^{\kappa \sigma }$. More specifically, the critical
current at $\sigma =9 kbar$ was found to be three times higher its
value at $\sigma =1.5 kbar$, clearly indicating a
weak-links-mediated origin of the phenomenon. Hence, for small
enough $\sigma $ we can safely assume that (Sergeenkov 1999)
$t(\sigma )\simeq t(0)(1-\beta \sigma/\sigma_0)$ with $\sigma _0$
being some characteristic value (the parameter $\beta $ is related
to the so-called ultimate stress $\sigma _m$ as $\beta =\sigma
_0/\sigma _m$). As a result, we have the following two
stress-induced effects in Josephson contacts:

(I) amplitude modulation leading to the explicit stress dependence
of the zero-field energy
\begin{equation}
J({\bf T},\sigma )=J({\bf T},0)e^{\gamma \sigma/\sigma_0}
\end{equation}
with $\gamma =\beta t(0)/\xi$, and\\

(II) phase modulation leading to the explicit stress dependence of
the flux
\begin{equation}
\Phi ({\bf T},{\bf H},\sigma )={\bf H}wd({\bf T},\sigma )
\end{equation}
with
\begin{equation}
d({\bf T},\sigma )=2\lambda _{L}({\bf T})+t(0)(1-\beta
\sigma/\sigma_0 )
\end{equation}

Finally, in view of Eqs.(36)-(41), the temperature and field
dependence of the small single junction TEC reads (the initial
phase difference is conveniently fixed at $\varphi _0=\pi$):
\begin{equation}
\alpha ({\bf T},{\bf H})=\alpha ({\bf T},0)\left [1+F({\bf T},{\bf
H})\right ]+\epsilon ({\bf T},0)\frac{dF({\bf T},{\bf H})}{d{\bf
T}} \label{TEcoefficient}
\end{equation}
where
\begin{equation}
F({\bf T},{\bf H})=\left [\frac{\sin
\varphi}{\varphi}+\frac{\xi}{d({\bf T},0)} \left (\frac{\sin
\varphi}{\varphi}-\cos \varphi \right )\right ]
\end{equation}
with
\begin{equation}
\varphi({\bf T},{\bf H})=\frac{\pi \Phi ({\bf T},{\bf H},0)}{\Phi
_0}=\frac{{\bf H}}{{\bf H}_0({\bf T})}
\end{equation}
\begin{equation}
\alpha ({\bf T},0)=\frac{d\epsilon ({\bf T},0)}{d{\bf T}}
\end{equation}
and
\begin{equation}
\epsilon ({\bf T},0)=-\left (\frac{\Phi _0}{2\pi}\right )\left
(\frac{2\gamma}{V \sigma_0}\right )I_C({\bf T})
\end{equation}
Here, ${\bf H}_0({\bf T})=\Phi _0/\pi wd({\bf T},0)$ with $d({\bf
T},0)=2\lambda _{L}({\bf T})+t(0)$.
\begin{SCfigure}[][ht]
\includegraphics[width=65mm]{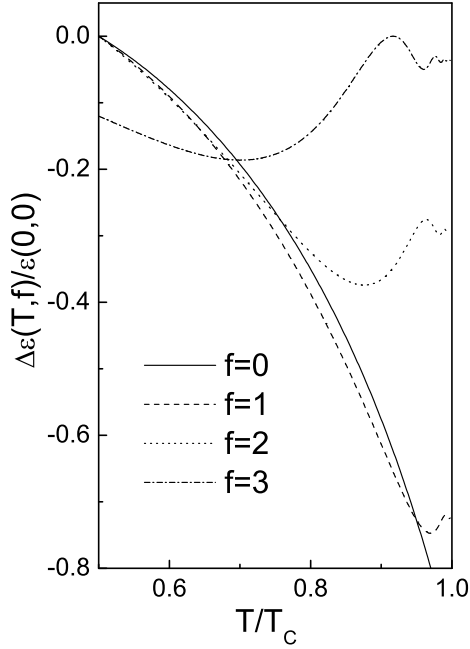}
\caption{Temperature dependence of the flux driven strain field in
a single short contact for different values of the frustration
parameter ${\bf f}$ according to Eqs.(36)-(48).}
 \label{fig13}
 \end{SCfigure}
For the explicit temperature dependence of $J({\bf T},0)=\Phi
_0I_C({\bf T})/2\pi$ we use the well-known (Meservey and Schwartz
1969, Sergeenkov 2002) analytical approximation of the BCS gap
parameter (valid for all temperatures), $\Delta ({\bf T})=\Delta
(0)\tanh \left(2.2\sqrt{\frac{{\bf T_C}-{\bf T}}{{\bf T}}}\right)$
with $\Delta (0)=1.76k_B{\bf T_C}$ which governs the temperature
dependence of the Josephson critical current
\begin{equation}
I_C({\bf T})=I_C(0)\left[ \frac{\Delta ({\bf T})}{\Delta
(0)}\right] \tanh \left[ \frac{\Delta ({\bf T})}{2k_{B}{\bf
T}}\right]
\end{equation}
while the temperature dependence of the London penetration depth
is governed by the two-fluid model:
\begin{equation}
\lambda _{L}({\bf T})=\frac{\lambda _{L}(0)}{\sqrt{1-({\bf T}/{\bf
T_C})^2}}
\end{equation}
\begin{SCfigure}[][ht]
\includegraphics[width=65mm]{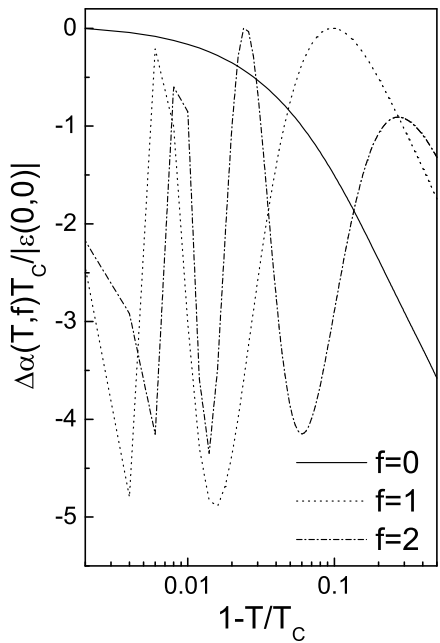}
\caption{Temperature dependence of flux driven normalized TEC in a
single small contact for different values of the frustration
parameter ${\bf f}$ (for the same set of parameters as in Fig.13)
according to Eqs.(36)-(48).}
 \label{fig14}
 \end{SCfigure}
From the very structure of Eqs.(36)-(44) it is obvious that TEC of
a single contact will exhibit {\it field} oscillations imposed by
the Fraunhofer dependence of the critical current $I_C$. Much less
obvious is its temperature dependence. Indeed, Fig.\ref{fig13}
presents the temperature behavior of the contact area strain field
$\Delta \epsilon ({\bf T},{\bf f})=\epsilon ({\bf T},{\bf
f})-\epsilon ({\bf T},0)$ (with $t(0)/\xi = 1$, $\xi /\lambda
_L(0)=0.02$ and $\beta =0.1$) for different values of the
frustration parameter ${\bf f}={\bf H}/{\bf H}_0(0)$. Notice
characteristic flux driven temperature oscillations near ${\bf
T_C}$ which are better seen on a semi-log plot shown in
Fig.\ref{fig14} which depicts the dependence of the properly
normalized field-induced TEC $\Delta \alpha ({\bf T},{\bf
f})=\alpha ({\bf T},{\bf f})-\alpha ({\bf T},0)$ as a function of
${\bf 1}-{\bf T}/{\bf T_C}$ for the same set of parameters.

To answer an important question how the neglected in the previous
analysis screening effects will affect the above-predicted
oscillating behavior of the field-induced TEC, let us consider a
more realistic situation with a junction embedded into an array
(rather than an isolated contact) which is realized in
artificially prepared arrays using photolithographic technique
that nowadays allows for controlled manipulations of the junctions
parameters (Newrock et al. 2000). Besides, this is also a good
approximation for a granular superconductor (if we consider it as
a network of superconducting islands connected with each other via
Josephson links). Our goal is to model and simulate the elastic
response of such systems to an effective stress $\sigma$. For
simplicity, we will consider an array with a regular topology and
uniform parameters (such approximation already proved useful for
describing high-quality artificially prepared structures, see,
e.g., Sergeenkov and Araujo-Moreira 2004).

Let us consider a planar square array as shown in Fig.\ref{fig15}.
The total current includes the bias current flowing through the
vertical junctions and the induced screening currents circulating
in the plaquette (Nakajima and Sawada 1981). This situation
corresponds to the inclusion of screening currents only into the
nearest neighbors, neglecting thus the mutual inductance terms
(Phillips et al. 1993). Therefore, the equation for the vertical
contacts will read (horizontal and vertical junctions are denoted
by superscripts $h$ and $v$, respectively):
\begin{figure}[ht]
\centerline{\includegraphics[width=85mm]{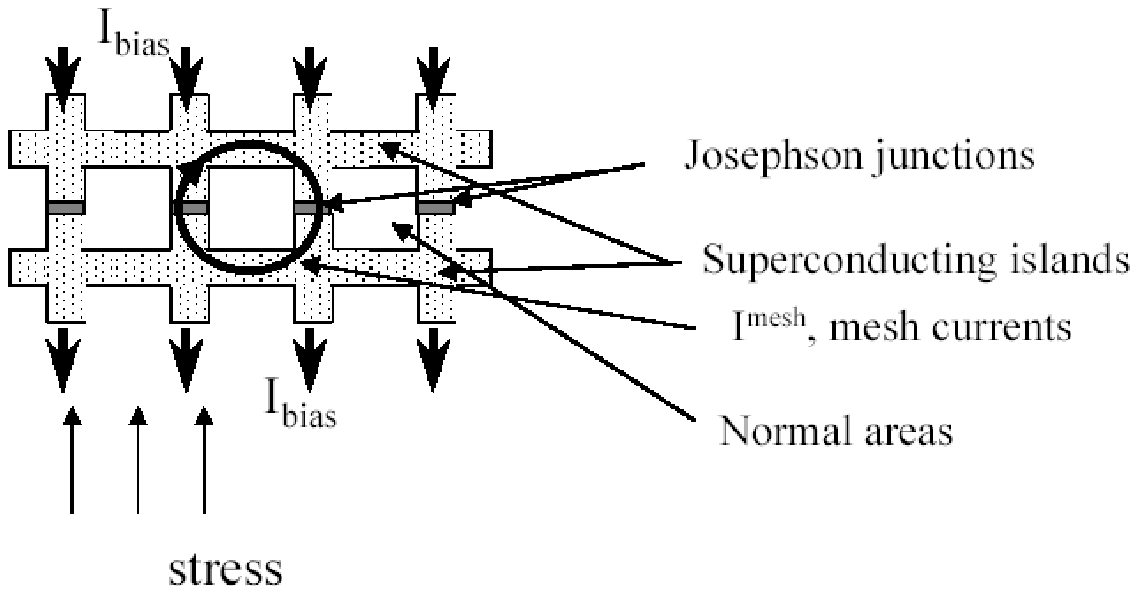} \hspace{0mm}
\includegraphics[width=80mm]{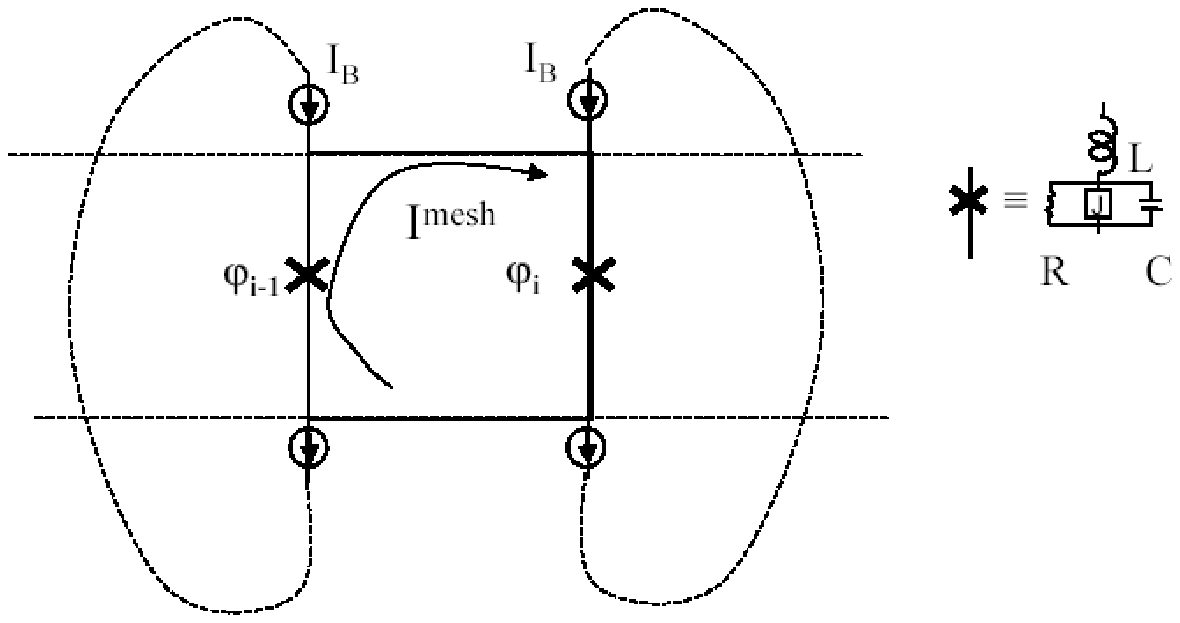}}
\caption{Left: sketch of a regular square array (a single
plaquette).  Right: electrical scheme of the array with the
circulating currents. The bias current is fed via virtual loops
external to the array.}
 \label{fig15}
 \end{figure}
\begin{equation}
\frac{\hbar C}{2e}\frac{d^2\phi_{i,j}^v}{dt^2} + \frac{\hbar}{2eR}
\frac{d\phi_{i,j}^v}{dt} +
 I_c \sin \phi_{i,j}^v = \Delta I^s_{i,j} + I_b
\label{currentcons-nostress}
\end{equation}
where $\Delta I^s_{i,j}=I^s_{i,j}-I^s_{i-1,j}$ and the screening
currents $I^s$ obey the fluxoid conservation condition:
\begin{equation}
-\phi^v_{i,j}+\phi^v_{i,j+1} - \phi^h_{i,j} + \phi^h_{i+1,j} =
2\pi \frac{\Phi^{ext}}{\Phi_0} - \frac{2\pi L I^s_{i,j}}{\Phi_0}
\label{fluxoid}
\end{equation}
Recall that the total flux has two components (an external
contribution and the contribution due to the screening currents in
the closed loop) and it is equal to the sum of the phase
differences describing the array. It is important to underline
that the external flux in Eq.(\ref{fluxoid}), $\eta = 2\pi
\Phi^{ext}/\Phi_ 0$, is related to the frustration of the whole
array, i.e., this is the flux across the void of the network
(Araujo-Moreira et al. 1997, Araujo-Moreira et al. 2005, Grimaldi
et al. 1996), and it should be distinguished from the previously
introduced applied magnetic field ${\bf H}$ across the junction
barrier which is related to the frustration of a single contact
${\bf f}=2\pi {\bf H}dw/\Phi _0$ and which only modulates the
critical current $I_C({\bf T},{\bf H},\sigma )$ of a single
junction while inducing a negligible flux into the void area of
the array.

For simplicity, in what follows we will consider only the elastic
effects due to a uniform (homogeneous) stress imposed on the
array. With regard to the geometry of the array, the deformation
of the loop is the dominant effect with its radius $a$ deforming
as follows:
\begin{equation}
a(\sigma)=a_0(1-\chi \sigma/\sigma_0)
\end{equation}
As a result, the self-inductance of the loop $L(a)=\mu_0 a F(a)$
(with $F(a)$ being a geometry dependent factor) will change
accordingly:
\begin{equation}
L(a)=L_0(1-\chi_g \sigma/\sigma_0) \label{induct-dep}
\end{equation}
The relationship between the coefficients $\chi$ and $\chi_g$ is
given by
\begin{equation}
\chi_g = \left( 1+a_0 B_g \right)\chi \label{chirelation}
\end{equation}
where $B_g=\frac{1}{F(a)}\left(\frac{dF}{da}\right)_{a_0}$.
\begin{SCfigure}[][ht]
\includegraphics[width=65mm]{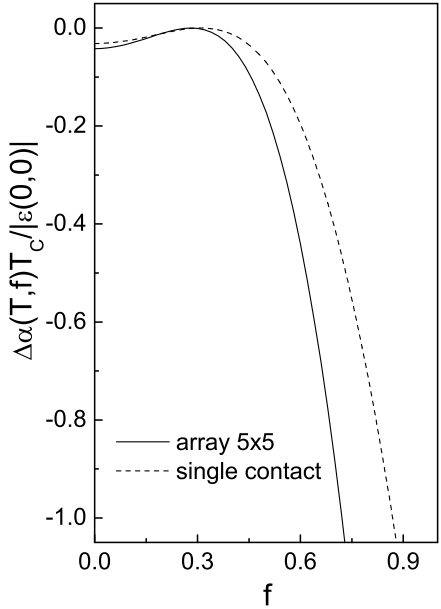}
\caption{Numerical simulation results for an array $5 \times 5$
(solid line) and a small single contact (dashed line). The
dependence of the normalized TEC  on the frustration parameter
${\bf f}$ (applied magnetic field ${\bf H}$ across the barrier)
for the reduced temperature ${\bf T}/{\bf T_C}=0.95$. The
parameters used for the simulations: $\eta =0$, $\beta = 0.1$,
$t(0)/\xi = 1$, $\xi/\lambda_L=0.02$, $\beta_L=10$,
$\gamma_b=0.95$, and $\chi_g=\chi=0.01$.}
 \label{fig16}
 \end{SCfigure}
It is also reasonable to assume that in addition to the critical
current, the external stress will modify the resistance of the
contact:
\begin{equation}
R(\sigma ) =\frac{\pi \Delta(0)}{2 e I_C(\sigma )}=R_0 e^{-\chi
\sigma/\sigma_0} \label{resistance}
\end{equation}
as well as capacitance (due to the change in the distance between
the superconductors):
\begin{equation}
C(\sigma )=\frac{C_0}{1-\chi \sigma/\sigma_0}\simeq C_0
(1+\chi\sigma/\sigma_0) \label{capacitance}
\end{equation}
To simplify the treatment of the dynamic equations of the array,
it is convenient to introduce the standard normalization
parameters such as the Josephson frequency:
\begin{equation}
\omega_J = \sqrt{\frac{2\pi I_C(0)}{C_0 \Phi_0}} \label{omegaj}
\end{equation}
the analog of the SQUID parameter:
\begin{equation}
\beta_L = \frac{2\pi I_C(0)L_0}{\Phi_0}, \label{squidpar}
\end{equation}
and the dissipation parameter:
\begin{equation}
\beta_C = \frac{2\pi I_C(0)C_0R_0^2}{\Phi_0} \label{betac}
\end{equation}
Combining Eqs.(\ref{currentcons-nostress}) and (\ref{fluxoid})
with the stress-induced effects described by Eqs.
(\ref{resistance}) and (\ref{capacitance}) and using the
normalization parameters given by
Eqs.(\ref{omegaj})-(\ref{betac}), we can rewrite the equations for
an array in a rather compact form. Namely, the equations for
vertical junctions read:
\begin{eqnarray}
\frac{1}{1-\chi \sigma/\sigma_0} \ddot{\phi}_{i,j}^v +
\frac{e^{-\chi \sigma/\sigma_0}}{\sqrt{\beta_C}}
\dot{\phi}_{i,j}^v +
e^{\chi \sigma/\sigma_0} \sin \phi_{i,j}^v  = \gamma_b + \nonumber \\
\frac{1}{\beta_L \left( 1 -\chi_g \sigma/\sigma_0 \right) } \left[
\phi^v_{i,j-1} - 2\phi^v_{i,j}  +  \phi^v_{i,j+1} + \phi^h_{i,j} -
\phi^h_{i-1,j} + \phi^h_{i+1,j-1} - \phi^h_{i,j-1}  \right]
\label{arr-eq-vert}
\end{eqnarray}
Here an overdot denotes the time derivative with respect to the
normalized time (inverse Josephson frequency), and the bias
current is normalized to the critical current without stress,
$\gamma_b = I_b/I_C(0)$.

The equations for the horizontal junctions will have the same
structure safe for the explicit bias related terms:
\begin{eqnarray}
\frac{1}{1-\chi \sigma/\sigma_0} \ddot{\phi}_{i,j}^h +
\frac{e^{-\chi \sigma/\sigma_0}}{\sqrt{\beta_C}}
\dot{\phi}_{i,j}^h +
e^{\chi \sigma/\sigma_0} \sin \phi_{i,j}^h =  \nonumber \\
\frac{1}{\beta_L \left( 1 -\chi_g \sigma/\sigma_0 \right) } \left[
\phi^h_{i,j-1} -2\phi^h_{i,j}+\phi^h_{i,j+1} + \phi^v_{i,j} -
\phi^v_{i-1,j} + \phi^v_{i+1,j-1} - \phi^v_{i,j-1} \right]
\label{arr-eq-hor}
\end{eqnarray}
Finally, Eqs.(\ref{arr-eq-vert}) and (\ref{arr-eq-hor}) should be
complemented with the appropriate boundary conditions (Binder et
al. 2000)  which will include the normalized contribution of the
external flux through the plaquette area $\eta = 2\pi
\Phi^{ext}/\Phi_0$.
\begin{figure}[ht]
\begin{center}
\includegraphics[width=65mm,height=8.250cm]{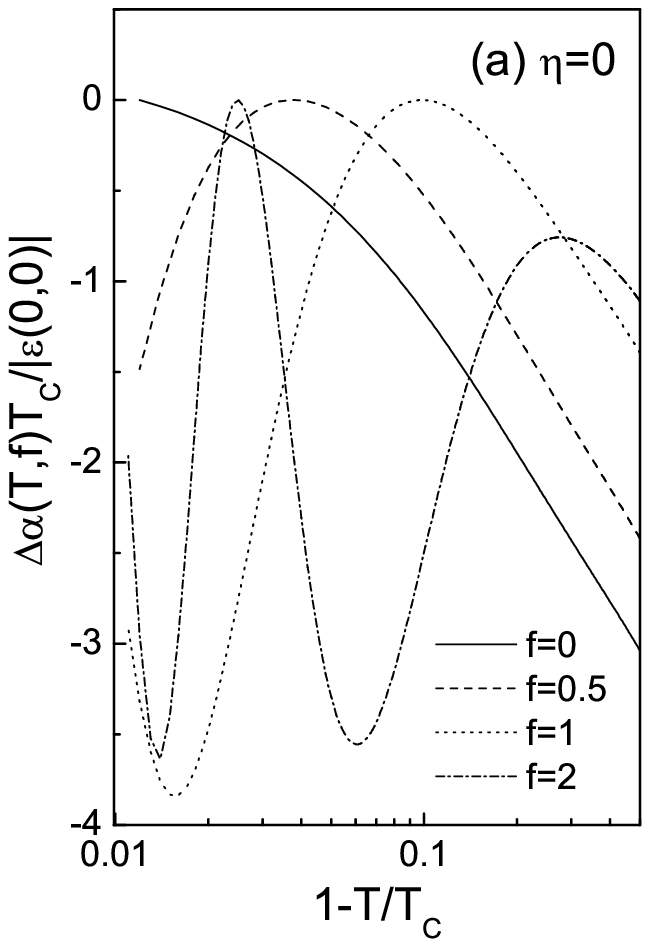}
\includegraphics[width=65mm,height=8.250cm]{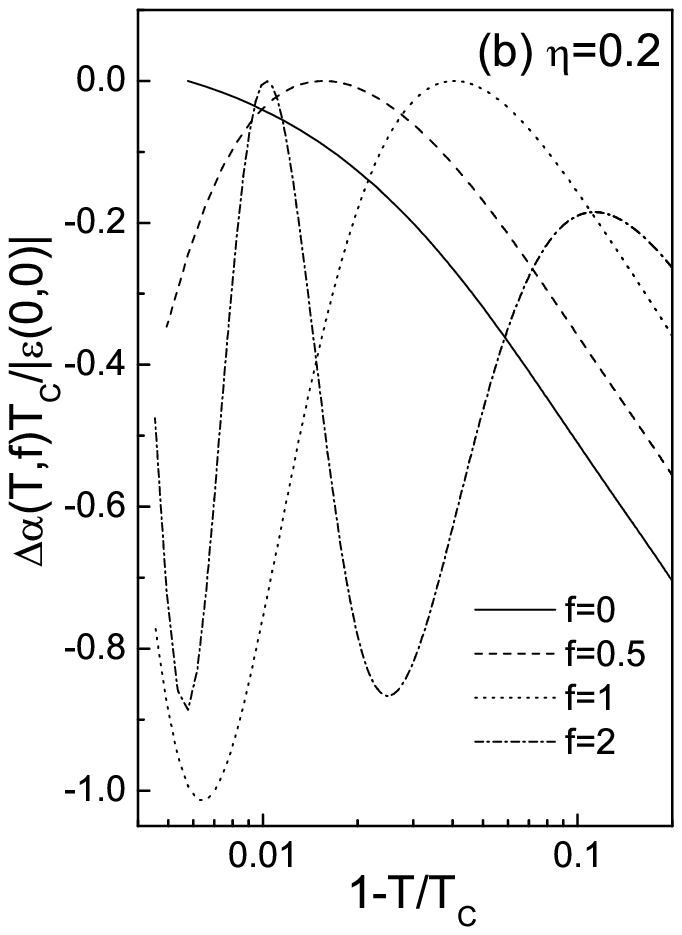}\\\vspace*{7mm}
\includegraphics[width=65mm,height=8.250cm]{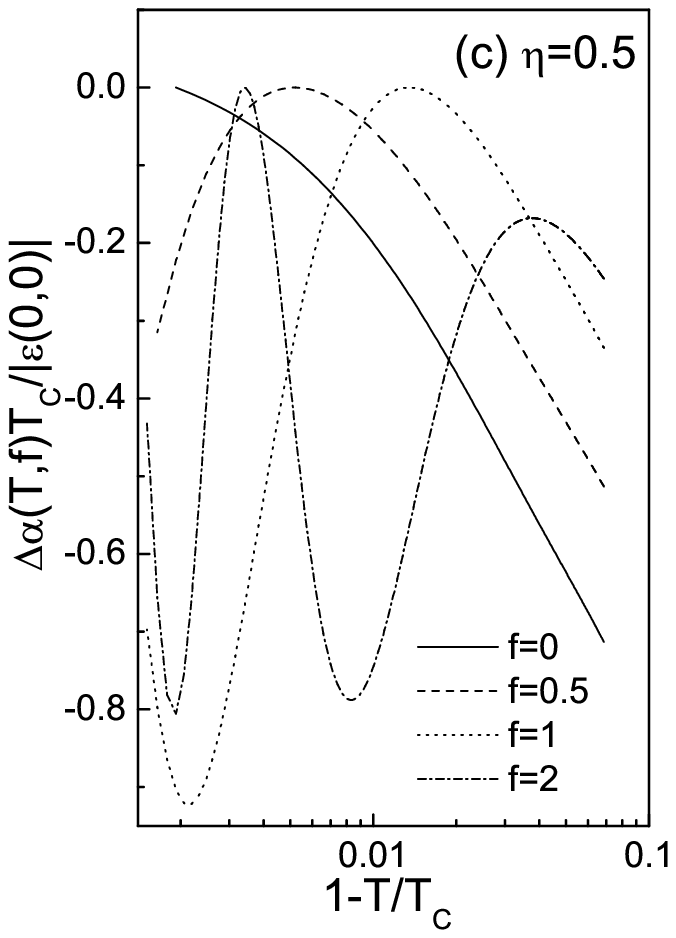}
\includegraphics[width=65mm,height=8.250cm]{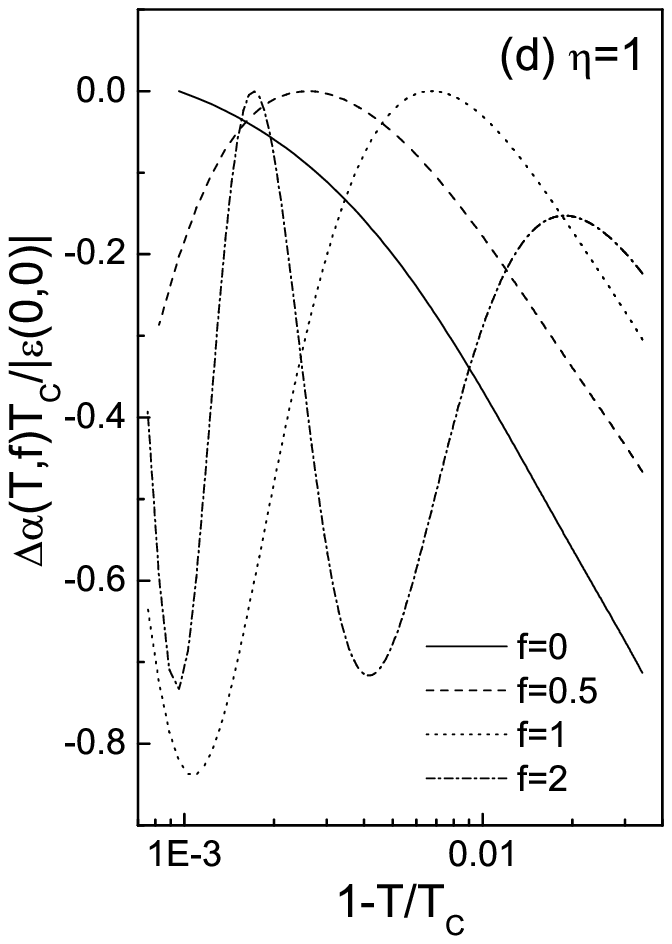}\end{center} \vspace*{-7mm}
\caption{Numerical simulation results for an array $5 \times 5$.
The influence of the flux across the void of the network $\eta$
frustrating the whole array on the temperature dependence of the
normalized TEC for different values of the barrier field ${\bf f}$
frustrating a single junction for $\gamma_b=0.5$ and the rest of
parameters same as in Fig.\ref{fig16}.}
 \label{fig17}
 \end{figure}
It is interesting to notice that Eqs.(\ref{arr-eq-vert}) and
(\ref{arr-eq-hor}) will have the same form as their stress-free
counterparts if we introduce the stress-dependent renormalization
of the parameters:
\begin{equation}
\tilde{\omega}_J = \omega_J e^{\chi \sigma /2\sigma_0}
\label{omegajren}
\end{equation}

\begin{equation}
\tilde{ \beta}_C = \beta_C e^{- 3 \chi \sigma/\sigma_0}
\label{betacren}
\end{equation}

\begin{equation}
\tilde{\beta}_L = \beta_L (1-\chi_g \sigma/\sigma_0) e^{\chi
\sigma/\sigma_0 } \label{squidparren}
\end{equation}

\begin{equation}
\tilde{\eta} = \eta (1- 2 \chi \sigma/\sigma_0 ) \label{etaren}
\end{equation}

\begin{equation}
\tilde{\gamma}_b = \gamma_b e^{-\chi \sigma/\sigma_0 }
 \label{gammaren}
\end{equation}
Turning to the discussion of the obtained numerical simulation
results, it should be stressed that the main problem in dealing
with an array is that the total current through the junction
should be retrieved by solving self-consistently the array
equations in the presence of screening currents. Recall that the
Josephson energy of a single junction for an arbitrary current $I$
through the contact reads:
\begin{equation}
E_J({\bf T},{\bf f},I)=E_J({\bf T},{\bf
f},I_C)\left[1-\sqrt{1-\left(\frac{I}{I_C}\right)^2} \right]
\label{Josenergybias}
\end{equation}
The important consequence of Eq.(\ref{Josenergybias}) is that if
no current flows in the array's junction, such junction will not
contribute to the TEC (simply because a junction disconnected from
the current generator will not contribute to the energy of the
system).

Below we sketch the main steps of the numerical procedure used to
simulate the stress-induced effects in the array:
\begin{itemize}
\item[(1)] a bias point $I_b$ is selected for the whole array;
\item[(2)] the parameters of the array (screening, Josephson frequency,
dissipation, etc) are selected and modified according to the
intensity of the applied stress $\sigma$;
\item[(3)] the array equations are simulated to retrieve the static configuration of
the phase differences for the parameters selected in step $2$;
\item[(4)] the total current flowing through the individual junctions is
retrieved as:
\begin{equation}
I^{v,h}_{i,j} = I_C\sin \phi^{v,h}_{i,j} \label{totalcurrent}
\end{equation}
\item[(5)] the energy dependence upon stress is numerically estimated using
the value of the total current $I^{v,h}_{i,j}$ (which is not
necessarily identical for all junctions) found in step $4$ via
Eq.(\ref{totalcurrent});
\item[(6)] the array energy $E_J^A$ is obtained by summing up the contributions of
all junctions with the above-found phase differences
$\phi^{v,h}_{i,j}$;
\item[(7)] the stress-modified screening currents $I^s_{i,j}({\bf T},{\bf H},\sigma )$
are computed using Eq.(\ref{fluxoid}) and inserted into the
magnetic energy of the array $E_M^A = \frac{1}{2L}
\Sigma_{i,j}(I^s_{i,j})^2$;
\item[(8)] the resulting strain field and TE coefficient of the array are
computed using numerical derivatives based on the finite
differences:
\begin{equation}
\epsilon^A \simeq \frac{1}{V} \left[
\frac{\Delta\left(E_M^A+E_J^A\right)}{\Delta
\sigma}\right]_{\Delta\sigma \rightarrow 0}
 \label{estimstrain}
\end{equation}

\begin{equation}
\alpha ({\bf T},{\bf H}) \simeq \frac{ \Delta\epsilon^A}{ \Delta
{\bf T}} \label{estimalpha}
\end{equation}
\end{itemize}
The numerical simulation results show that the overall behavior of
the strain field and TE coefficient in the array is qualitatively
similar to the behavior of the single contact. In Fig.\ref{fig16}
we have simulated the behavior of both the small junction and the
array as a function of the field across the barrier of the
individual junctions in the presence of bias and screening
currents. As is seen, the dependence of $\alpha ({\bf T},{\bf f})$
is very weak up to ${\bf f}\simeq 0.5$, showing a strong decrease
of about $50 \%$ when the frustration approaches ${\bf f}=1$.

A much more profound change is obtained by varying the temperature
for the fixed value of applied magnetic field. Fig.\ref{fig17}
depicts the temperature behavior of $\alpha ({\bf T},{\bf f})$ (on
semi-log scale) for different field configurations which include
barrier field $f$ frustrating a single junction and the flux
across the void of the network $\eta$ frustrating the whole array.
First of all, comparing Fig.\ref{fig17}(a) and Fig.\ref{fig14} we
notice that, due to substantial modulation of the Josephson
critical current $I_C({\bf T},{\bf H})$ given by
Eq.(\ref{Josenergy}), the barrier field ${\bf f}$ has similar
effects on the TE coefficient of both the array and the single
contact including temperature oscillations. However, finite
screening effects in the array result in the appearance of
oscillations at higher values of the frustration ${\bf f}$ (in
comparison with a single contact). On the other hand,
Fig.\ref{fig17}(b-d) represent the influence of the external field
across the void $\eta$ on the evolution of $\alpha ({\bf T},{\bf
f})$. As is seen, in comparison with a field-free configuration
(shown in Fig.\ref{fig17}(a)), the presence of external field
$\eta$ substantially reduces the magnitude of the TE coefficient
of the array. Besides, with $\eta$ increasing, the onset of
temperature oscillations markedly shifts closer to ${\bf T_C}$.

\vspace{4mm} \leftline{ 6. SUMMARY} \vspace{5mm}

In this Chapter, using a realistic model of 2D Josephson junction
arrays (created by 2D network of twin boundary dislocations with
strain fields acting as an insulating barrier between hole-rich
domains in underdoped crystals), we considered many novel effects
related to the magnetic, electric, elastic and transport
properties of Josephson nanocontacts and nanogranular
superconductors. Some of the topics covered here include such
interesting phenomena as chemomagnetism and magnetoelectricity,
electric analog of the "fishtai" anomaly and field-tuned weakening
of the chemically-induced Coulomb blockade as well as a giant
enhancement of nonlinear thermal conductivity (reaching $500\%$
when the intrinsically induced chemoelectric field $E_\mu \propto
|\nabla \mu |$, created by the gradient of the chemical potential
due to segregation of hole producing oxygen vacancies, closely
matches the externally produced thermoelectric field $E_T \propto
|\nabla T|$). Besides, we have investigated the influence of a
homogeneous mechanical stress on a small single Josephson junction
and on a plaquette (array of $5\times 5$ junctions) and have shown
how the stress-induced modulation of the parameters describing the
junctions (as well as the connecting circuits) produces such an
interesting phenomenon as a thermal expansion (TE) in a single
contact and two-dimensional array (plaquette). We also studied the
variation of the TE coefficient with an external magnetic field
and temperature. In particular, near ${\bf T_C}$ (due to some
tremendous increase of the effective "sandwich" thickness of the
contact) the field-induced TE coefficient of a small junction
exhibits clear {\it temperature} oscillations scaled with the
number of flux quanta crossing the contact area. Our numerical
simulations revealed that these oscillations may actually still
survive in an array if the applied field is strong enough to
compensate for finite screening induced self-field effects.

The accurate estimates of the model parameters suggest quite an
optimistic possibility to experimentally realize all of the
predicted in this Chapter promising and important for applications
effects in non-stoichiometric nanogranular superconductors and
artificially prepared arrays of Josephson nanocontacts.

\vspace{8mm} \leftline{ACKNOWLEDGMENTS} \vspace{5mm}

Some of the results presented in Section 5 were obtained in
collaboration with Giacomo Rotoli and Giovanni Filatrella. This
work was supported by the Brazilian agency CAPES.

\newpage {\leftline {REFERENCES}} \vspace{5mm}

\noindent Abrikosov, A.A. (1988) {\it Fundamentals of the Theory
of Metals}, Elsevier, Amsterdam.

\noindent Akopyan, A.A., Bolgov, S.S. and  Savchenko, A.P. (1990)
Sov. Phys. Semicond. 24 1167.

\noindent Altshuler, E. and  Johansen, T.H. (2004) Rev. Mod. Phys.
76 471.

\noindent Anshukova, N.V.,  Bulychev, B.M., Golovashkin, A.I.,
Ivanova, L.I., Minakov,A.A. and Rusakov, A.P. (2000) JETP Lett. 71
377.

\noindent Araujo-Moreira, F.M., Barbara, P., Cawthorne, A.B. and
Lobb, C.J. (1997) Phys. Rev. Lett. 78 4625.

\noindent Araujo-Moreira, F.M., Barbara, P., Cawthorne, A.B. and
Lobb, C.J. (2002) {\em Studies of High Temperature
Superconductors} 43, (Ed. Narlikar, A.V.) Nova Science Publishers,
New York, p. 227.

\noindent Araujo-Moreira, F.M., Maluf, W. and Sergeenkov, S.
(2004) Solid State Commun. 131 759.

\noindent Araujo-Moreira, F.M., Maluf, W. and Sergeenkov, S.
(2005) Eur. Phys. J. B 44 33.

\noindent Barbara, P., Araujo-Moreira, F.M., Cawthorne, A.B. and
Lobb, C.J. (1999) Phys. Rev. B 60 7489.

\noindent Beloborodov, I.S., Lopatin, A.V.,  Vinokur, V.M. and
Efetov, K.B. (2007) Rev. Mod. Phys. 79 469.

\noindent Binder, P., Caputo, P.,  Fistul, M.V.,  Ustinov, A.V.
and Filatrella, G. (2000) Phys. Rev. B 62 8679.

\noindent Bourgeois, O,  Skipetrov, S.E., Ong, F. and Chaussy, J.
(2005) Phys. Rev. Lett. 94  057007.

\noindent Chen, W.,  Smith, T.P., Buttiker, M. et al. (1994) Phys.
Rev. Lett. 73 146.

\noindent Daeumling, M., Seuntjens, J.M. and Larbalestier, D.C.
(1990) Nature 346 332.

\noindent De Leo, C. and Rotoli, G. (2002) Phys. Rev. Lett. 89
167001.

\noindent Deppe, J. and  Feldman, J.L. (1994) Phys. Rev. B 50
6479.

\noindent D'yachenko, A.I.,  Tarenkov, V.Y., Abalioshev, A.V.,
Lutciv, L.V., Myasoedov, Y.N. and Boiko,Y.V. (1995) Physica C 251
207.

\noindent Eichenberger, A.-L., Affolter, J., Willemin, M.,
Mombelli, M., Beck, H., Martinoli,P.  and Korshunov, S.E. (1996)
Phys. Rev. Lett. 77 3905.

\noindent Gantmakher, V.F.,  Neminskii, A.M. and  Shovkun, D.V.
(1990) JETP Lett. 52 630.

\noindent Gantmakher, V.F. (2002) Physics-Uspekhi 45 1165.

\noindent Geim, A.K.,  Dubonos, S.V.,  Lok, J.G.S. et al. (1998)
Nature 396 144.

\noindent Girifalco, L.A. (1973) {\it Statistical Physics of
Materials}, A Wiley-Interscience, New York.

\noindent Golubov, A.A.,  Kupriyanov, M.Yu. and  Fominov, Ya.V.
(2002) JETP Lett. 75 588.

\noindent Grimaldi, G., Filatrella, G.,  Pace, S. and Gambardella,
U. (1996) Phys. Lett. A 223 463.

\noindent Gurevich, A. and  Pashitskii, E.A. (1997) Phys. Rev. B
56
 6213.

\noindent Guttman, G.,  Nathanson, B., Ben-Jacob, E. and Bergman,
D.J. (1997) Phys. Rev. B 55 12691.

\noindent Haviland, D.B., Kuzmin, L.S. and  Delsing, P. (1991) Z.
Phys. B 85 339.

\noindent Iansity, M.,  Johnson, A.J. and Lobb, C.J. (1988) Phys.
Rev. Lett. 60 2414.

\noindent Khaikin, M.S. and Khlyustikov, I.N. (1981) JETP Lett. 33
158.

\noindent Krive, I.V., Kulinich, S.I. and Jonson, M. (2004) Low
Temp. Phys. 30 554.

\noindent Lang, K.M., Madhavan, V., Hoffman, J.E., Hudson, E.W.,
Eisaki, H., Uchida, S. and Davis, J.C. (2002) Nature 415 412.

\noindent Li, M.S. (2003) Phys. Rep. 376 133.

\noindent Makhlin, Yu., Schon, G. and Shnirman, A. (2001) Rev.
Mod. Phys. 73 357.

\noindent Meservey, R. and Schwartz, B.B. (1969) {\em
Superconductivity}, vol.1, (Ed. Parks, R.D.), M. Dekker, New York,
p.117.

\noindent Moeckley, B.H., Lathrop, D.K. and  Buhrman, R.A.
 (1993) Phys. Rev. B 47 400.

\noindent Nagel, P., Pasler, V., Meingast, C.,  Rykov, R.I. and
Tajima, S. (2000) Phys. Rev. Lett. 85 2376.

\noindent Nakajima, K. and Sawada, Y. (1981) J. Appl. Phys. 52
5732.

\noindent Newrock, R.S., Lobb, C.J., Geigenmuller, U. and Octavio,
M. (2000) Solid State Phys. 54 263.

\noindent Orlando, T.P. and Delin, K.A. (1991) {\it Foundations of
Applied Superconductivity}, Addison, New York.

\noindent Ostrovsky, P.M. and  Feigel'man, M.V. (2004) JETP Lett.
79 489.

\noindent Phillips, J.R.,  van der Zant, R.S.J. and Orlando, T.P.
(1993) Phys. Rev. B 47 5219.

\noindent Ryazanov, V.V.,  Oboznov, V.A. and  Rusanov, A.Yu.
(2001) Phys. Rev. Lett. 86 2427.

\noindent Sergeenkov, S. and Ausloos, M. (1993) Phys. Rev. B 48
604.

\noindent Sergeenkov, S. (1995) J. Appl. Phys. 78 1114.

\noindent Sergeenkov, S. (1997) J. de Physique I (France) 7 1175.

\noindent Sergeenkov, S. (1998a) JETP Lett. 67 680.

\noindent Sergeenkov, S. (1998b) J. Phys.: Condensed Matter 10
L265.

\noindent Sergeenkov, S. (1999) JETP Lett. 70 36.

\noindent Sergeenkov, S. and Ausloos, M. (1999) JETP 89 140.

\noindent Sergeenkov, S. (2001) {\em Studies of High Temperature
Superconductors} 39, (Ed. Narlikar, A.V.) Nova Science Publishers,
New York, p. 117.

\noindent Sergeenkov, S. (2002) JETP Lett. 76 170.

\noindent Sergeenkov, S. (2003) JETP Lett. 77 94.

\noindent Sergeenkov, S. and Araujo-Moreira, F.M. (2004) JETP
Lett. 80 580.

\noindent Sergeenkov, S. (2005) JETP  101 919.

\noindent Sergeenkov, S. (2006) {\it Studies of High Temperature
Superconductors} 50, (Ed. Narlikar, A.V.) Nova Science Publishers,
New York, p. 229.

\noindent Sergeenkov, S. (2007) J. Appl. Phys. 102 066104.

\noindent Sergeenkov, S., Rotoli, G., Filatrella, G. and
Araujo-Moreira, F.M. (2007) Phys. Rev. B 75 014506.

\noindent van Bentum, P.J.M.,  van Kempen, H.,  van de Leemput,
L.E.C. and Teunissen, P.A.A. (1988) Phys. Rev. Lett. 60 369.

\noindent van der Zant, H.S.J. (1996) Physica B222 344.

\noindent van Harlingen, D.,  Heidel, D.F. and  Garland, J.C.
(1980) Phys. Rev. B 21 1842.

\noindent Wendin, G. and Shumeiko, V.S. (2007) Low Temp. Phys. 33
724.

\noindent Yang, G., Shang, P.,  Sutton, S.D.,  Jones, I.P., Abell,
J.S. and Gough, C.E. (1993) Phys. Rev. B 48
 4054.

\end{document}